\documentclass[12pt]{article}

\usepackage{booktabs}
\usepackage{amsthm,amsmath,amssymb,authblk}
\usepackage{algorithm} 
\usepackage{algorithmic}
\usepackage{color,soul}
\usepackage{graphicx}

\usepackage{caption}
\newtheorem{theorem}{Theorem}        
\newtheorem{proposition}{Proposition}

\newtheorem{condition}{Condition}    

\usepackage{lmodern} 
\usepackage[]{natbib}
\usepackage{bookmark}

\newcommand{\anon}{1}

\begin{document}

\def\spacingset#1{\renewcommand{\baselinestretch}%
{#1}\small\normalsize} \spacingset{1}

\if1\anon
{  
	\title{\bf Cross-Semantic Transfer Learning for High-Dimensional Linear Regression}
	
	
	\author[1]{Jiancheng Jiang\textsuperscript{$\dagger$}}
	
	\author[2]{Xuejun Jiang\textsuperscript{*}}
	
	\author[2]{Hongxia Jin\textsuperscript{$\dagger$}}
	
	\affil[1]{{\it{Department of Mathematics, Great Bay University, Dongguan, China}}}
	\affil[2]{{\it{Department of Statistics and Data Science, Southern University of Science and Technology, Shenzhen, China}}}
	
	\date{}      
	\maketitle
	
	\begingroup
	
	\renewcommand\thefootnote{$\dagger$} 
	\footnotetext{These authors contributed equally to this work.}
	
	\renewcommand\thefootnote{*} 
	\footnotetext{Corresponding author. Email address: jiangxj@sustech.edu.cn.}
	
	\endgroup
} \fi

\if0\anon
{
  \bigskip
  \bigskip
  \bigskip
  \begin{center}
    {\LARGE\bf Cross-Semantic Transfer Learning for High-Dimensional Linear Regression}
\end{center}
  \medskip
} \fi

\bigskip
\begin{abstract}
Current transfer learning methods for high-dimensional linear regression assume feature alignment across domains, restricting their applicability to semantically matched features. In many real-world scenarios, however, distinct features in the target and source domains can play similar predictive roles, creating a form of cross-semantic similarity. To leverage this broader transferability, we propose the Cross-Semantic Transfer Learning (CSTL) framework. It captures potential relationships by comparing each target coefficient with all source coefficients through a weighted fusion penalty. The weights are derived from the derivative of the SCAD penalty, effectively approximating an ideal weighting scheme that preserves transferable signals while filtering out source-specific noise. For computational efficiency, we implement CSTL using the Alternating Direction Method of Multipliers (ADMM). Theoretically, we establish that under mild conditions, CSTL achieves the oracle estimator with overwhelming probability. Empirical results from simulations and a real-data application confirm that CSTL outperforms existing methods in both cross-semantic and partial signal similarity settings.
\end{abstract}

\noindent%
{\it Keywords:} Transfer Learning, Cross-semantic Signal Similarity, Fusion Penalty, SCAD
\vfill

\newpage
\spacingset{1.8} 

\section{Introduction}\label{sec1}

Transfer learning, which enhances performance in data-scarce target domains by leveraging knowledge from related, data-rich source domains \citep{pan2009survey,torrey2010transfer}, has achieved remarkable success in fields like computer vision \citep{yosinski2014transferable,zoph2018learning,kornblith2019better} and natural language processing \citep{devlin2019bert,raffel2020exploring}. For comprehensive overviews, see \citet{weiss2016survey} and \citet{zhuang2020comprehensive}.

The high-dimensional regression setting, where the number of covariates exceeds the sample size, presents significant challenges for estimation and prediction. This has spurred a growing body of research on transfer learning for high-dimensional data. Seminal works by \citet{bastani2021predicting} and \citet{li2022transfer} introduced two-step frameworks for single-source and multi-source scenarios, respectively. Subsequent studies have built on this idea by incorporating fusion penalties on the differences between source and target coefficients to enable more selective information transfer \citep{gao2023transfer,liu2024unified,he2024transfusion}. 
Beyond linear regression, these methodologies have been extended to a wide range of models, including generalized linear models \citep{tian2023transfer,li2024estimation}, Gaussian graphical models \citep{li2023transfer}, functional linear regression \citep{lin2022transfer}, quantile regression \citep{zhang2022transfer,jin2024transfer}, and nonparametric regression \citep{cai2024transfer}. 
A common thread among these diverse approaches is their reliance on an assumption of global similarity between domains -- an assumption that is often too restrictive in practice. To overcome this limitation, recent research has focused on frameworks for partial information transfer, where only a subset of parameters is shared. Representative works in this vein include \citet{he2024adatrans} and \citet{zhang2024covariate}.

While prior studies predominantly focus on homogeneous feature spaces—where source and target domains share the same set of covariates—many real-world applications involve heterogeneous covariate sets. This heterogeneity commonly arises in multi-source integration, cross-platform studies, or federated analysis, and has recently attracted growing research interest. For instance, \citet{zhao2023heterogeneous} examine settings where source-domain covariates constitute a subset of those in the target domain, whereas \citet{chang2024heterogeneous} investigate the opposite scenario where the target domain observes only a subset of source features.

A common thread across both homogeneous and heterogeneous transfer learning literature is the assumption that certain covariates must be semantically aligned across domains, with transfer restricted to their corresponding coefficients. This requirement, however, substantially limits the applicability of transfer learning. In practice, covariates with different semantics may still play similar roles, leading to comparable coefficient patterns. For example, in medical studies, body mass index (BMI) in the target domain and waist-to-hip ratio in a source domain differ semantically and in measurement, yet both reflect obesity-related information and can exhibit similar coefficients when predicting cardiovascular disease risk. We refer to this phenomenon as cross-semantic signal similarity.

This observation implies that even in the absence of explicit semantic alignment, source domains may still contain valuable information for target tasks. It thus reveals a significant opportunity for cross-semantic signal transfer, which existing alignment-dependent methods are unable to capture.

Motivated by this observation, we adopt a novel perspective: instead of comparing coefficients only across semantically aligned features, we compare each target coefficient with all source-domain coefficients. This all-pairs approach enables the complete identification of transferable signals, thereby unlocking richer information from the source domain. Our main contributions are summarized as follows:

\begin{itemize}
	\item[(i)] A Novel Framework and Efficient Algorithm:
	We propose a new transfer learning framework, termed ``Cross-Semantic Transfer Learning'' (CSTL),
	based on a weighted fusion penalty applied to all possible target-source coefficient pairs. This design allows CSTL to selectively borrow information from the source domain while filtering out non-transferable signals, leading to improved estimation and prediction in the target task. We also develop an efficient optimization algorithm based on the Alternating Direction Method of Multipliers (ADMM) to implement the proposed method.
	\item[(ii)] Theoretical Guarantees with Data-Driven Weights:
	We first establish the oracle property of CSTL in an ideal scenario where the true model structure is known, demonstrating that it achieves the oracle estimator with ideal weights. For practical implementation, we propose data-driven weights constructed from the derivative of the SCAD penalty to approximate this ideal weighting scheme. Under mild regularity conditions, we prove that the resulting CSTL estimator consistently identifies the true transferable structure and attains oracle performance with overwhelming probability.
	\item[(iii)] Empirical Validation:
	Through comprehensive simulations and a real-data application, we demonstrate that CSTL effectively captures transferable signals under both cross-semantic and partial similarity settings. The results show that our method outperforms existing alternatives and achieves performance comparable to the oracle benchmark.
\end{itemize}

The remainder of the paper is organized as follows. Section 2 introduces the problem setup and key notation. Section 3 introduces the proposed CSTL framework. It details the construction of adaptive weights and presents several numerical results in low-dimensional settings, which exemplify the advantages of CSTL.
Theoretical guarantees for the CSTL estimator are established in Section 4. Section 5 describes the efficient optimization algorithm for implementation. Finally, Section 6 validates the method's performance through comprehensive simulation studies and a real-data application.

\section{Preliminaries}
\subsection{Problem Setup}
We study transfer learning for high-dimensional linear regression, aiming to improve the estimation accuracy of target-domain coefficients by leveraging auxiliary data from a related, data-rich source domain. The target domain data 
$\{ (\mathbf{X}^{(t)}_i, Y_i^{(t)}) \}_{i=1}^{n_t} $
consist of a design matrix
$\mathbf{X}^{(t)} \in \mathbb{R}^{n_t \times d_t} $ and a response vector $ \mathbf{Y}^{(t)} \in \mathbb{R}^{n_t}$. Similarly, the source domain data $\{ (\mathbf{X}^{(s)}_i, Y_i^{(s)}) \}_{i=1}^{n_s}$
consist of $\mathbf{X}^{(s)} \in \mathbb{R}^{n_s \times d_s} $ and $ \mathbf{Y}^{(s)} \in \mathbb{R}^{n_s} $. These data are assumed to follow the linear models:
\begin{equation}\label{eq:dgp}
	\mathbf{Y}^{(t)} = \mathbf{X}^{(t)} \boldsymbol{\beta}^{*} + \boldsymbol{\epsilon}^{(t)}, \qquad
	\mathbf{Y}^{(s)}= \mathbf{X}^{(s)} \boldsymbol{\theta}^{*} + \boldsymbol{\epsilon}^{(s)},
\end{equation}
where $\boldsymbol{\beta}^* \in \mathbb{R}^{d_t}$ and $\boldsymbol{\theta}^* \in \mathbb{R}^{d_s}$ are the true sparse coefficient vectors, and $\boldsymbol{\epsilon}^{(t)}$ and $\boldsymbol{\epsilon}^{(s)}$ are independent noise vectors.

Departing from existing methods that require semantic alignment of covariates, we introduce a more flexible notion of transferability. Specifically, a source coefficient $\theta_l^*$
is considered transferable if there exists a target coefficient
$\beta_j^*$ such that $\beta_j^* $ equals $\theta_l^*$, irrespective of whether their corresponding covariates share the same semantic meaning. In the following we formalize this cross-semantic transferability.

\subsection{Notation}
To formalize coefficient-level transferability in high-dimensional regression, we first introduce the transfer structure set
\begin{equation}\label{eq:def-B}
	\mathcal{B} = \left\{ (j, l) \in [d_t] \times [d_s] : \beta_j^{*} = \theta_l^{*} \right\},
\end{equation}
where $[q]:= \{1,2,\dots,q\}$ for any positive integer $q$.  This set collects all target–source coefficient pairs that have the same value, regardless of semantic correspondence of the features. Let
\begin{equation}\label{eq:def-A}
	\mathcal{A}_t = \{ j \in [d_t] : \beta_j^{*} \ne 0 \}\quad\text{and} \quad \mathcal{A}_s = \{ l \in [d_s] : \theta_l^{*} \ne 0 \}
\end{equation}
denote the supports of the target and source models, respectively. We further decompose them into shared (transferable) and domain-specific (non-transferable) parts:
\begin{align*}
	\mathcal{T}_t & = \left\{ j \in \mathcal{A}_t : \exists\, l \in \mathcal{A}_s \text{ s.t. } \beta_j^{*} = \theta_l^{*} \right\},\quad 
	\mathcal{I}_t = \mathcal{A}_t \setminus \mathcal{T}_t,\\
	\mathcal{T}_s & = \left\{ l \in \mathcal{A}_s : \exists\, j \in \mathcal{A}_t \text{ s.t. } \theta_l^{*} = \beta_j^{*} \right\}, \quad
	\mathcal{I}_s = \mathcal{A}_s \setminus \mathcal{T}_s.
\end{align*}
Here, $\mathcal{T}_t$ and $\mathcal{T}_s$ denote the index sets of coefficients that are shared across the two domains (and thus transferable), while $\mathcal{I}_t$ and $\mathcal{I}_s$ denote the index sets of domain-specific coefficients. 

To clearly represent the relationships among coefficient values shared across domains, including one-to-one, many-to-one, and many-to-many correspondences, 
we define the canonical representative set
\begin{equation}\label{eq:def-tTt}
	\tilde{\mathcal{T}}_t = \left\{ \min \left\{ j \in \mathcal{T}_t : \beta_j^{*} = v \right\} : v \in \left\{ \beta_j^{*} : j \in \mathcal{T}_t \right\} \right\},
\end{equation} 
which selects a unique index for each distinct shared coefficient value. Let $m = |\tilde{\mathcal{T}}_t|$ be the number of distinct shared values, and denote the corresponding values by $\boldsymbol{\beta}_{\tilde{\mathcal{T}}_t}^{*} = (\alpha_1^{*}, \dots, \alpha_m^{*})^\top$. We then introduce binary matching matrices $\mathbf{M}^{(t)} \in \{0,1\}^{|\mathcal{T}_t| \times m}$ and $\mathbf{M}^{(s)} \in \{0,1\}^{|\mathcal{T}_s| \times m}$, defined by
\begin{equation}\label{eq:M}
	[\mathbf{M}^{(t)}]_{i,r} = \mathbb{I}\{ \beta^{*}_{j_i} = \alpha_r^{*} \}, \quad 
	[\mathbf{M}^{(s)}]_{i,r} = \mathbb{I}\{ \theta^{*}_{l_i} = \alpha_r^{*} \},
\end{equation}  
where $j_i$ and $l_i$ denote the $i$-th smallest index in $\mathcal{T}_t$ and $\mathcal{T}_s$, respectively. 

With these constructs, the shared coefficients in both domains can be compactly expressed as
\begin{equation*}
	\boldsymbol{\beta}_{\mathcal{T}_t}^{*} = \mathbf{M}^{(t)} \boldsymbol{\beta}_{\tilde{\mathcal{T}}_t}^{*}, \quad
	\boldsymbol{\theta}_{\mathcal{T}_s}^{*} = \mathbf{M}^{(s)} \boldsymbol{\beta}_{\tilde{\mathcal{T}}_t}^{*}.
\end{equation*}

\begin{figure}[!htbp]
	\centering
	\includegraphics[width=0.9\textwidth]{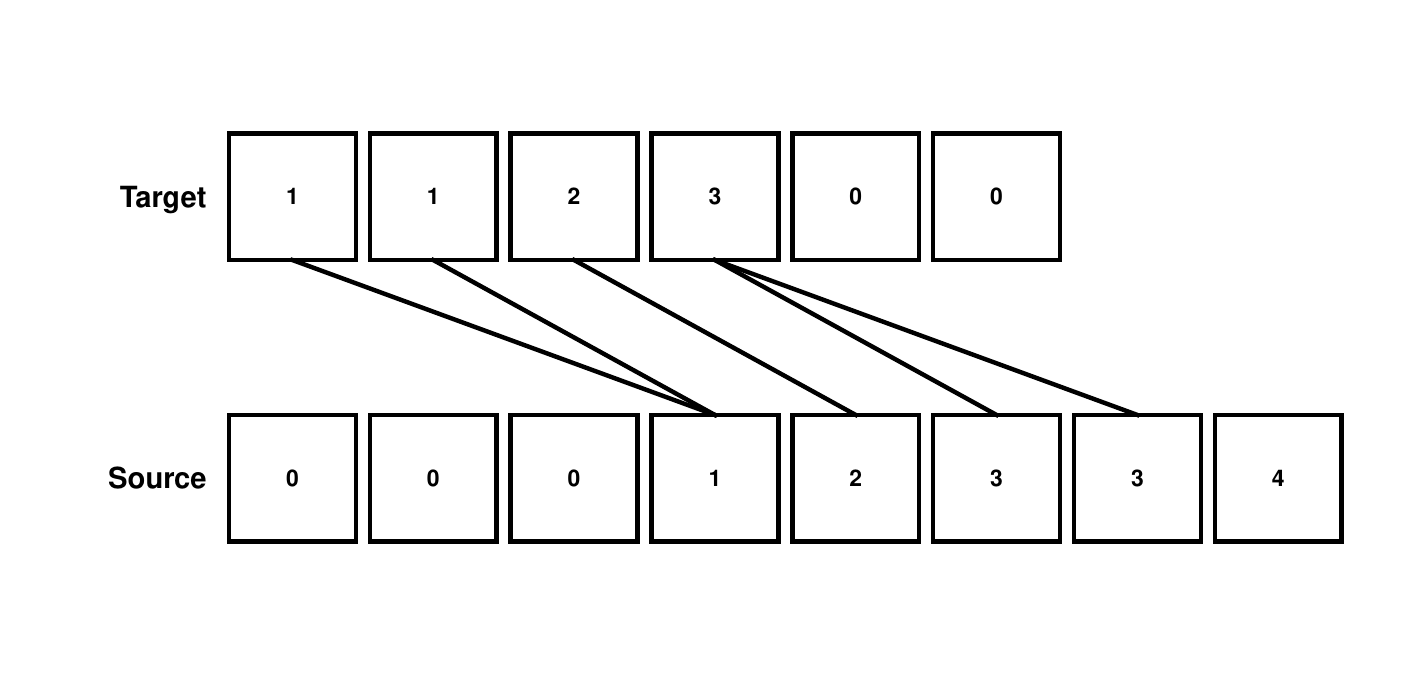}
	\vspace{-1.8em}
	\caption{  \label{toy figure}
		\textrm{Illustration of cross-domain signal sharing:
			(i) Blocks - Represent coefficients, arranged by feature index (target domain on top, source on bottom).
			(ii) Numbers - Indicate the true coefficient values. 
			(iii) Solid Lines - Each line connects a pair of coefficients in the set $\mathcal{B}$,
			indicating they share the same value across domains.}
	}
\end{figure} 

To illustrate the above constructions, we consider the following toy example.

\noindent\textbf{Toy Example.}
Consider the target and source coefficient vectors shown in Figure~\ref{toy figure}:
\begin{equation*}
	\boldsymbol{\beta}^{*} = (1,1,2,3,0,0)^\top,
	\qquad
	\boldsymbol{\theta}^{*} = (0,0,0,1,2,3,3,4)^\top.
\end{equation*} 
Then the supports are $\mathcal{A}_t=\{1,2,3,4\}$ and $\mathcal{A}_s=\{4,5,6,7,8\}$. The shared (transferable) sets are $\mathcal{T}_t = \{1,2,3,4\}$ and $\mathcal{T}_s = \{4,5,6,7\}$, consisting exactly of the indices that are connected by solid lines in Figure~\ref{toy figure}, while the remaining indices are domain-specific: $\mathcal{I}_t = \emptyset$ and $\mathcal{I}_s = \{8\}$.
The canonical representative set is $\tilde{\mathcal{T}}_t=\{1,3,4\}$, 
corresponding to distinct transferable values $\boldsymbol{\beta}_{\tilde{\mathcal{T}}_t}^{*}  = (1,2,3)$. Accordingly, the matching matrices are
{\renewcommand{\arraystretch}{0.85}
\begin{equation*}
	\mathbf{M}^{(t)}=
	\begin{bmatrix}
		1 & 0 & 0 \\
		1 & 0 & 0 \\
		0 & 1 & 0 \\
		0 & 0 & 1
	\end{bmatrix},
	\qquad
	\mathbf{M}^{(s)}=
	\begin{bmatrix}
		1 & 0 & 0 \\
		0 & 1 & 0 \\
		0 & 0 & 1 \\
		0 & 0 & 1
	\end{bmatrix}.
\end{equation*} 
}

\section{Method}
We aim to recover the sparse signal in the target model by fully leveraging auxiliary information from the source domain, even when transferable signals are connected via cross-semantic similarity rather than direct feature alignment. This is achieved by extending coefficient comparison to all target-source pairs and encouraging fusion for those with similar values. Formally, we propose the Cross-Semantic Transfer Learning (CSTL) framework, where the estimator is derived by solving the following regularized problem:
\begin{equation}\label{original}
	\min _{\boldsymbol{\beta}, \boldsymbol{\theta}}\Bigl\{
	\frac{1}{n_t} \left\|\mathbf{Y}^{(t)} - \mathbf{X}^{(t)} \boldsymbol{\beta} \right\|_2^2
	+ \frac{1}{n_s} \left\|\mathbf{Y}^{(s)} - \mathbf{X}^{(s)} \boldsymbol{\theta} \right\|_2^2
	+\lambda_0 \sum_{j=1}^{d_t} w_{j} \left| \beta_j \right|
	+ \lambda_1 \sum_{j=1}^{d_t} \sum_{l=1}^{d_s} w_{j,l} \left| \beta_j - \theta_l \right|
	\Bigr\},
\end{equation}
where $\lambda_0$ and $\lambda_1$ are tuning parameters, and \( w_j \) and \( w_{j,l} \) are adaptive weights. 
At first glance, one might suppose that an $L_1$-penalty on $\boldsymbol{\theta}$ for the objective function in \eqref{original} is needed to induce sparsity when $\theta_{\ell}=0$. In fact, such a penalty is unnecessary, as it is already implicitly enforced via the fourth term when $\beta_j=0$.
However, if all $\beta_j$'s are nonzero, the fourth term provides no direct penalty on $\theta_{\ell}$. In that scenario, the same sparsity effect can still be realized by introducing a noise variable into the target model whose corresponding true coefficient is zero.

The objective function in (\ref{original}) consists of four parts. The first two terms are the sample-size–normalized empirical losses for the target and source domains, which prevents the large-sample source domain from dominating the objective function. The third term imposes an $\ell_1$ penalty on the target coefficients to enforce sparsity, while the final term penalizes pairwise differences $|\beta_j - \theta_l|$ over all target–source coefficient pairs, thereby enforcing full pairwise fusion between the target and source domains. This formulation allows transferable information to be discovered directly at the coefficient level, without assuming covariate alignment across domains. Crucially, the strength of each pairwise penalty is governed by a weight $w_{j,l}$: coefficient pairs with similar values receive larger weights that encourage fusion, while heterogeneous pairs receive smaller weights and remain separate. The optimization problem in \eqref{original} can be efficiently solved using the ADMM algorithm; see Section \ref{admm} for details.

The above CSTL framework is designed for two linear regression models, but it can be extended to transfer knowledge between any two supervised learning models. This is achieved by replacing the residual sum of squares with other appropriate loss functions, thereby allowing shared coefficient pairs to be fused and improving estimation efficiency. 

The implementation of CSTL requires appropriate weight specification. We begin our analysis with an ideal setting where the target sparsity pattern and transferable structure are known.

\subsection{Ideal Weights under Known Transfer Structures}\label{sec:idealw}
Suppose the target support set \( \mathcal{A}_t \subseteq [d_t] \) and the
transferable structure set \( \mathcal{B} \subseteq [d_t] \times [d_s] \) were known.
In this ideal setting, we design the weights in \eqref{original} to penalize only non-active target coefficients and transferable coefficient pairs:
\begin{equation}\label{eq:idealw}
	w_j = \mathbb{I}(j \in \mathcal{A}_t^c)
	\quad \mbox{\rm and}\quad
	w_{j,l} = \mathbb{I}((j,l) \in \mathcal{B}),
\end{equation} 
or equivalently
$
w_j=\mathbb I(\beta_j^*=0)$ and
$w_{j,\ell}=\mathbb I(\beta_j^*=\theta_\ell^*).
$
With such ideal weights, taking $\lambda_0, \lambda_1 \to \infty$ enforces the constraints 
$\beta_j = 0$ for $j \in \mathcal{A}_t^c$ and $\beta_j = \theta_l$ for $(j,l)\in\mathcal{B}$. 
In this limit, the penalized estimator coincides with the oracle estimator 
$(\hat{\boldsymbol{\beta}}_{\mathrm{ora}}, \hat{\boldsymbol{\theta}}_{\mathrm{ora}})$, defined as the solution to:
\begin{equation}\label{oral}
	\begin{array}{cl}
		\min _{\boldsymbol{\beta}, \boldsymbol{\theta}} & \left\{\frac{1}{n_t}  \left\|\mathbf{Y}^{(t)}-\mathbf{X}^{(t)}\boldsymbol{\beta}\right\|_2^2+\frac{1}{n_s}\left\|\mathbf{Y}^{(s)}-\mathbf{X}^{(s)} \boldsymbol{\theta}\right\|_2^2\right\} \\
		\text { s.t. }                                             & \boldsymbol{\beta}_{\mathcal{A}_t^{c}}=0, \forall (j, l)\in \mathcal{B}, \beta_j = \theta_l.
	\end{array}
\end{equation}

The constraints in (\ref{oral}) imply that 
\begin{equation*}
	\boldsymbol{\theta}_{\mathcal{A}_s^{c}}=0, \quad 
	\boldsymbol{\beta}_{\mathcal{T}_t} = \mathbf{M}^{(t)} \boldsymbol{\beta}_{\tilde{\mathcal{T}}_t}, \quad \mbox{\rm and}\quad
	\boldsymbol{\theta}_{\mathcal{T}_s} = \mathbf{M}^{(s)} \boldsymbol{\beta}_{\tilde{\mathcal{T}}_t},
\end{equation*}
where \( \tilde{\mathcal{T}}_t \), \( \mathbf{M}^{(t)} \), and \( \mathbf{M}^{(s)} \) are as defined in \eqref{eq:def-tTt} and \eqref{eq:M} (see Appendix for details of the reconstruction). With the structural constraints imposed, the oracle estimator reduces to
\begin{equation}\label{oralvariant}
	\begin{aligned}
		\hat{\boldsymbol{\beta}}_{\mathrm{ora}},\;
		\hat{\boldsymbol{\theta}}_{\mathrm{ora}}
		\in
		\operatorname*{argmin}_{\substack{
				\boldsymbol{\beta}_{\mathcal{A}_t^{c}}=0,\;
				\boldsymbol{\beta}_{\mathcal{T}_t}=\mathbf{M}^{(t)}\boldsymbol{\beta}_{\tilde{\mathcal{T}}_t},\\[2pt]
				\boldsymbol{\theta}_{\mathcal{A}_s^{c}}=0,\;
				\boldsymbol{\theta}_{\mathcal{T}_s}=\mathbf{M}^{(s)}\boldsymbol{\beta}_{\tilde{\mathcal{T}}_t}
		}}
		&\Bigg\{
		\frac{1}{n_t}
		\left\|
		\mathbf{Y}^{(t)} -
		\big(
		\widetilde{\mathbf{X}}^{(t)}_{\tilde{\mathcal{T}}_t}\,\boldsymbol{\beta}_{\tilde{\mathcal{T}}_t}
		+ \mathbf{X}^{(t)}_{\mathcal{I}_t}\,\boldsymbol{\beta}_{\mathcal{I}_t}
		\big)
		\right\|_2^2
		\\
		&+ \frac{1}{n_s}
		\left\|
		\mathbf{Y}^{(s)} -
		\big(
		\widetilde{\mathbf{X}}^{(s)}_{\tilde{\mathcal{T}}_t}\,\boldsymbol{\beta}_{\tilde{\mathcal{T}}_t}
		+ \mathbf{X}^{(s)}_{\mathcal{I}_s}\,\boldsymbol{\theta}_{\mathcal{I}_s}
		\big)
		\right\|_2^2
		\Bigg\},
	\end{aligned}
\end{equation}
where $\tilde{\mathbf{X}}_{\tilde{\mathcal{T}}_t}^{(t)}=\mathbf{X}_{\mathcal{T}_t}^{(t)}\mathbf{M}^{(t)}$ and $\tilde{\mathbf{X}}_{\tilde{\mathcal{T}}_t}^{(s)}=\mathbf{X}_{\mathcal{T}_s}^{(s)}\mathbf{M}^{(s)}$. Under mild identifiability conditions, the estimator admits a closed formula, as indicated in Proposition~\ref{pro-oracle-closed}. 

To simplify the notation of the following result, we define
\begin{equation}\label{eq:pooled-design}
	\mathbf{Y}=
	\begin{pmatrix}
		\tfrac{1}{\sqrt{n_t}} \, \mathbf{Y}^{(t)} \\[6pt]
		\tfrac{1}{\sqrt{n_s}} \, \mathbf{Y}^{(s)}
	\end{pmatrix},
	\qquad
	\tilde{\mathbf{X}}_{\tilde{\mathcal{T}}_t}=
	\begin{pmatrix}
		\tfrac{1}{\sqrt{n_t}}
		\bigl(\mathbf{I} - \boldsymbol{P}_{\mathcal{I}_t}^{(t)}\bigr)
		\tilde{\mathbf{X}}_{\tilde{\mathcal{T}}_t}^{(t)} \\[6pt]
		\tfrac{1}{\sqrt{n_s}}
		\bigl(\mathbf{I} - \boldsymbol{P}_{\mathcal{I}_s}^{(s)}\bigr)
		\tilde{\mathbf{X}}_{\tilde{\mathcal{T}}_t}^{(s)}
	\end{pmatrix},
\end{equation}
where  $\boldsymbol{P}_{\mathcal{I}_k}^{(k)}=\mathbf{X}_{\mathcal{I}_k}^{(k)}\left[\left(\mathbf{X}_{\mathcal{I}_k}^{(k)}\right)^{\top} \mathbf{X}_{\mathcal{I}_k}^{(k)}\right]^{-1}\left(\mathbf{X}_{\mathcal{I}_k}^{(k)}\right)^{\top}$ is the projection matrix for $k\in\{t,s\}$.

\begin{proposition}\label{pro-oracle-closed}
	Suppose $\left|\tilde{\mathcal{T}}_t\right|<n_t$,$\left|\mathcal{I}_t\right|<n_t$ and  $\left|\mathcal{I}_s\right|<n_s$. Then the oracle estimators of (\ref{oral}) are given by
\end{proposition}
\vspace{-0.5cm}
\begin{equation}\label{tTtclosed}
	\begin{aligned}
		\hat{\boldsymbol{\beta}}_{\mathrm{ora},\tilde{\mathcal{T}}_t} 
		& = \left[\frac{1}{n_t}\left(\tilde{\mathbf{X}}_{\tilde{\mathcal{T}}_t}^{(t)}\right)^{\top} \left(\boldsymbol{I}-\boldsymbol{P}_{\mathcal{I}_t}^{(t)}\right)\tilde{\mathbf{X}}_{\tilde{\mathcal{T}}_t}^{(t)}  + \frac{1}{n_s}\left(\tilde{\mathbf{X}}_{\tilde{\mathcal{T}}_t}^{(s)}\right)^{\top}\left(\boldsymbol{I}-\boldsymbol{P}_{\mathcal{I}_s}^{(s)}\right)\tilde{\mathbf{X}}_{\tilde{\mathcal{T}}_t}^{(s)} \right]^{-1} \\
		& \quad \times \left[\frac{1}{n_t}\left(\tilde{\mathbf{X}}_{\tilde{\mathcal{T}}_t}^{(t)}\right)^{\top}\left(\boldsymbol{I}-\boldsymbol{P}_{\mathcal{I}_t}^{(t)}\right)\mathbf{Y}^{(t)} + \frac{1}{n_s}\left(\tilde{\mathbf{X}}_{\tilde{\mathcal{T}}_t}^{(s)}\right)^{\top}\left(\boldsymbol{I}-\boldsymbol{P}_{\mathcal{I}_s}^{(s)}\right)\mathbf{Y}^{(s)} \right]\\
		&=\left[\left(\tilde{\mathbf{X}}_{\tilde{\mathcal{T}}_t}\right)^{\top}\tilde{\mathbf{X}}_{\tilde{\mathcal{T}}_t}\right]^{-1}\left(\tilde{\mathbf{X}}_{\tilde{\mathcal{T}}_t}\right)^{\top} \mathbf{Y},
	\end{aligned}
\end{equation}

\begin{equation}\label{Itclosed}
	\hat{\boldsymbol{\beta}}_{\mathrm{ora},\mathcal{I}_t}=\left[\left(\mathbf{X}_{\mathcal{I}_t}^{(t)}\right)^{\top}\mathbf{X}_{\mathcal{I}_t}^{(t)} \right]^{-1}\left(\mathbf{X}_{\mathcal{I}_t}^{(t)}\right)^{\top}\left[\mathbf{Y}^{(t)}-\tilde{\mathbf{X}}_{\tilde{\mathcal{T}}_t}^{(t)}\hat{\boldsymbol{\beta}}_{\mathrm{ora},\tilde{\mathcal{T}}_t}\right] ,\quad \text{and} \quad \hat{\boldsymbol{\beta}}_{\mathrm{ora},\mathcal{A}_t^c}=\mathbf{0}.
\end{equation}

Proposition~\ref{pro-oracle-closed} shows that $\hat{\boldsymbol{\beta}}_{\mathrm{ora},\tilde{\mathcal{T}}_t}$ 
corresponds to the least-squares fit of the responses on the transferable covariates after removing the effects of domain-specific variables, and $\hat{\boldsymbol{\beta}}_{\mathrm{ora},\mathcal{I}_t}$ corresponds to regressing the remaining target residuals (after subtracting the contribution of the estimated transferable part) on the target-specific covariates. 

To better understand this closed-form expression, two special cases are considered:
\begin{itemize}
	\item[(1)] If for $k\in\{t,s\}$, the transferable covariates are orthogonal to the domain-specific ones (i.e., $\tilde{\mathbf{X}}_{\tilde{\mathcal{T}}_t}^{(k)} \perp \mathbf{X}_{\mathcal{I}_k}^{(k)}$), then $\tilde{\mathbf{X}}_{\tilde{\mathcal{T}}_t}=\left(\left(\tilde{\mathbf{X}}_{\tilde{\mathcal{T}}_t}^{(t)}\right)^{\top}/\sqrt{n_t},\left(\tilde{\mathbf{X}}_{\tilde{\mathcal{T}}_t}^{(s)}\right)^{\top}/\sqrt{n_s}\right)^{\top}$, implying that $\hat{\boldsymbol{\beta}}_{\mathrm{ora},\tilde{\mathcal{T}}_t}$ 
	is determined entirely by the transferable covariates.
	\item[(2)] If $\tilde{\mathcal{T}}_t=\emptyset$, no coefficients are transferable, and thus the source domain provides no useful information to the target. 
	In this case, the target component of the oracle estimator reduces to the ordinary least-squares solution:
	\begin{equation*}
		\hat{\boldsymbol{\beta}}_{\mathrm{ora},\mathcal{A}_t}
		= \hat{\boldsymbol{\beta}}_{\mathrm{ols},\mathcal{A}_t},\quad
		\hat{\boldsymbol{\beta}}_{\mathrm{ora},\mathcal{A}_t^c}=\mathbf{0}.
	\end{equation*}
\end{itemize}

The preceding analysis establishes the oracle properties of the CSTL estimator under an ideal weighting scheme. This theoretical ideal serves as a benchmark, motivating the practical, data-driven weight selection method developed in the following subsection.

\subsection{Data-driven Weights for Transferable Structure Recovery}\label{sec:adaweight}

If the true sparsity pattern and transferable structure were known, the ideal weights in \eqref{eq:idealw} could be expressed as
$w_{j} = \mathbb{I}(\beta_{j}^{*} = 0)$ and $w_{j, l} = \mathbb{I}(|\beta_{j}^{*} - \theta_{l}^{*}| = 0)$. In practice, however, these oracle quantities are unknown.
A natural approach is to approximate the ideal weights using initial estimators. For example, one could apply Lasso separately to the target and source data to obtain $\hat{\boldsymbol{\beta}}_{\text{init}}$ and $\hat{\boldsymbol{\theta}}_{\text{init}}$, and subsequently define
$\hat{w}_{j} = \mathbb{I}(\hat{\beta}_{\text{init},j} = 0), \quad
\hat{w}_{j,l} = \mathbb{I}(|\hat{\beta}_{\text{init},j} - \hat{\theta}_{\text{init},l}| = 0)$.
Unfortunately, this hard-thresholding rule is unstable under finite samples, largely due to the effects of shrinkage bias and estimation noise.

To address this instability, we implement a data-driven weighting scheme inspired by the adaptive fusion approach of \citet{he2024adatrans}. This method leverages the derivative of the SCAD penalty \citep{fan2001variable} to define the weights smoothly. Specifically, we set
\begin{equation*}
	\hat{w}_{j} = \frac{1}{\lambda_0} \cdot p'_{\lambda_0} \left( \left| \hat{\beta}_{\text{init},j} \right| \right), \quad \mbox{\rm and}\quad
	\hat{w}_{j,l} = \frac{1}{\lambda_1} \cdot p'_{\lambda_1} \left( \left|\hat{\beta}_{\text{init},j} - \hat{\theta}_{\text{init},l} \right| \right),
\end{equation*} 
where the SCAD derivative $p_\lambda'(t)$ is
\begin{equation*}
	p'_\lambda(t) =
	\begin{cases}
		\lambda \cdot \operatorname{sgn}(t),
		& \text{if } |t| \leq \lambda,\\
		\frac{\operatorname{sgn}(t)(a\lambda - |t|)}{a-1},      & \text{if } \lambda < |t| \leq a\lambda,
		\quad a>2,\\
		0, & \text{if } |t| > a\lambda.
	\end{cases}
\end{equation*} 

\begin{figure}[!htbp]
	\centering
	\includegraphics[width=0.7\textwidth]{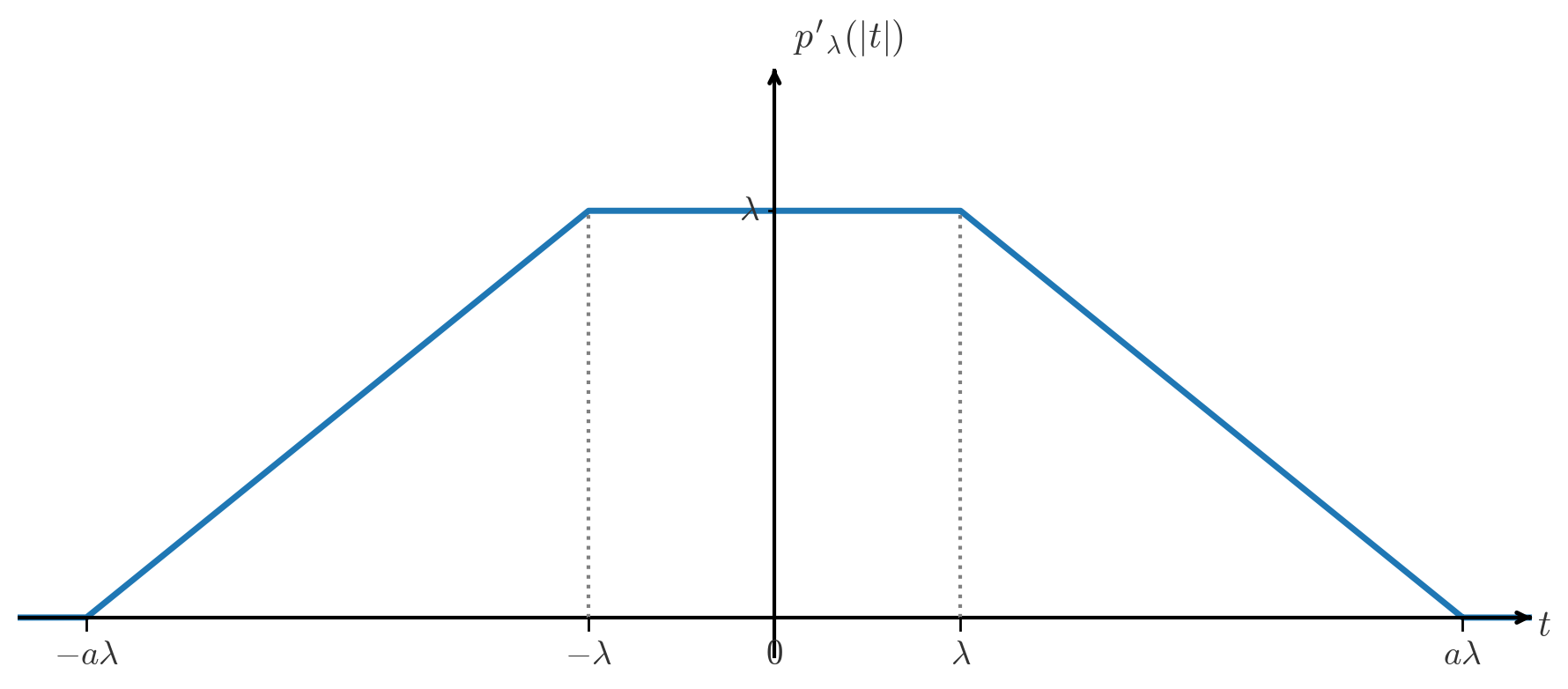}
	\caption{Derivative of the SCAD penalty function $p'_\lambda(|t|)$ with parameter $a>2$.}
	\label{fig:scad-derivative}
\end{figure} 

Figure~\ref{fig:scad-derivative} illustrates the behavior of $p'_{\lambda} (|t|)$, which acts as a smooth relaxation of the hard-thresholding rule. This design yields a principled weighting scheme: when coefficients or their differences are small, the weights are near one, applying strong penalties to encourage sparsity and fusion. For intermediate values, the weights decrease linearly, mitigating the instability inherent in hard-thresholding. Finally, for large values, the weights vanish entirely, ensuring that strong signals and significant cross-domain differences remain unpenalized. By smoothly interpolating between these regimes, the proposed weighting scheme closely approximates the ideal oracle weights, effectively enforcing sparsity on the target coefficients while distinguishing transferable signals from domain-specific ones in the source data. We summarize the complete CSTL procedure in Algorithm~\ref{alg:CSTL}, which follows the implementation framework of Algorithm 1 in \citet{he2024adatrans}.

\begin{algorithm}[H]
	\setlength{\baselineskip}{20pt} 
	\caption{Cross-Semantic Transfer Learning (CSTL)}
	\label{alg:CSTL}
	\textbf{Input:} Target data \( (\mathbf{X}^{(t)}, \mathbf{Y}^{(t)}) \);
	source data \( (\mathbf{X}^{(s)}, \mathbf{Y}^{(s)}) \);
	tuning parameters \( \lambda_0, \lambda_1 \). \\
	\textbf{Output:} Estimated target coefficients \( \hat{\boldsymbol{\beta}}_{\text{cst}} \). \\
	\textbf{Step 1.} Obtain initial estimators $\hat{\boldsymbol{\beta}}_{\text{init}}$ and $\hat{\boldsymbol{\theta}}_{\text{init}}$
	via Lasso regression separately to the target and source data. \\
	\textbf{Step 2.} Construct data-driven weights from the initial estimators:
	\[
	\hat{w}_{j} = \frac{1}{\lambda_0} \cdot
	p'_{\lambda_0}\!\left( \left| \hat{\beta}_{\text{init},j} \right| \right), \quad
	\hat{w}_{j,l} = \frac{1}{\lambda_1} \cdot
	p'_{\lambda_1}\!\left( \left| \hat{\beta}_{\text{init},j}
	- \hat{\theta}_{\text{init},l} \right| \right).
	\] 
	\textbf{Step 3.} Solve the following optimization problem to obtain
	\( \hat{\boldsymbol{\beta}}_{\text{cst}} \),
	\( \hat{\boldsymbol{\theta}}_{\text{cst}} \):
	\[
	\min_{\boldsymbol{\beta}, \boldsymbol{\theta}} \left\{
	\frac{1}{n_t} \left\| \mathbf{Y}^{(t)} - \mathbf{X}^{(t)} \boldsymbol{\beta} \right\|_2^2
	+ \frac{1}{n_s} \left\| \mathbf{Y}^{(s)} - \mathbf{X}^{(s)} \boldsymbol{\theta} \right\|_2^2
	+ \lambda_0 \sum_{j=1}^{d_t} \hat{w}_{j} \left| \beta_j \right|
	+ \lambda_1 \sum_{j=1}^{d_t} \sum_{l=1}^{d_s} \hat{w}_{j,l} \left| \beta_j - \theta_l \right|
	\right\}.
	\]
\end{algorithm}

\subsection{Low-dimensional Illustrations}

This subsection presents two low-dimensional examples to (i) illustrate how the oracle estimator can be computed explicitly when the equality set $\mathcal B$ is known, and (ii) examine whether the data-driven CSTL estimator successfully identifies the transferable pairs when $\mathcal B$ is unknown. High-dimensional numerical results are presented in Section \ref{sec:simulation}.

Throughout this subsection, we consider the linear models in \eqref{eq:dgp}.
For the purpose of these low-dimensional illustrations, we generate covariates and noise
from simple Gaussian distributions: for $k\in\{t,s\}$, the rows of
$\mathbf X^{(k)}$ are i.i.d.\ $N(\mathbf 0,\mathbf I_3)$ and
$\boldsymbol{\epsilon}^{(k)}\sim N(\mathbf 0,\mathbf I_{n_k})$.
We set $n_t=100$, $n_s=200$, and repeat each experiment 500 times.

\begin{figure}[htbp]  
	\centering  
	\includegraphics[width=0.90\textwidth]{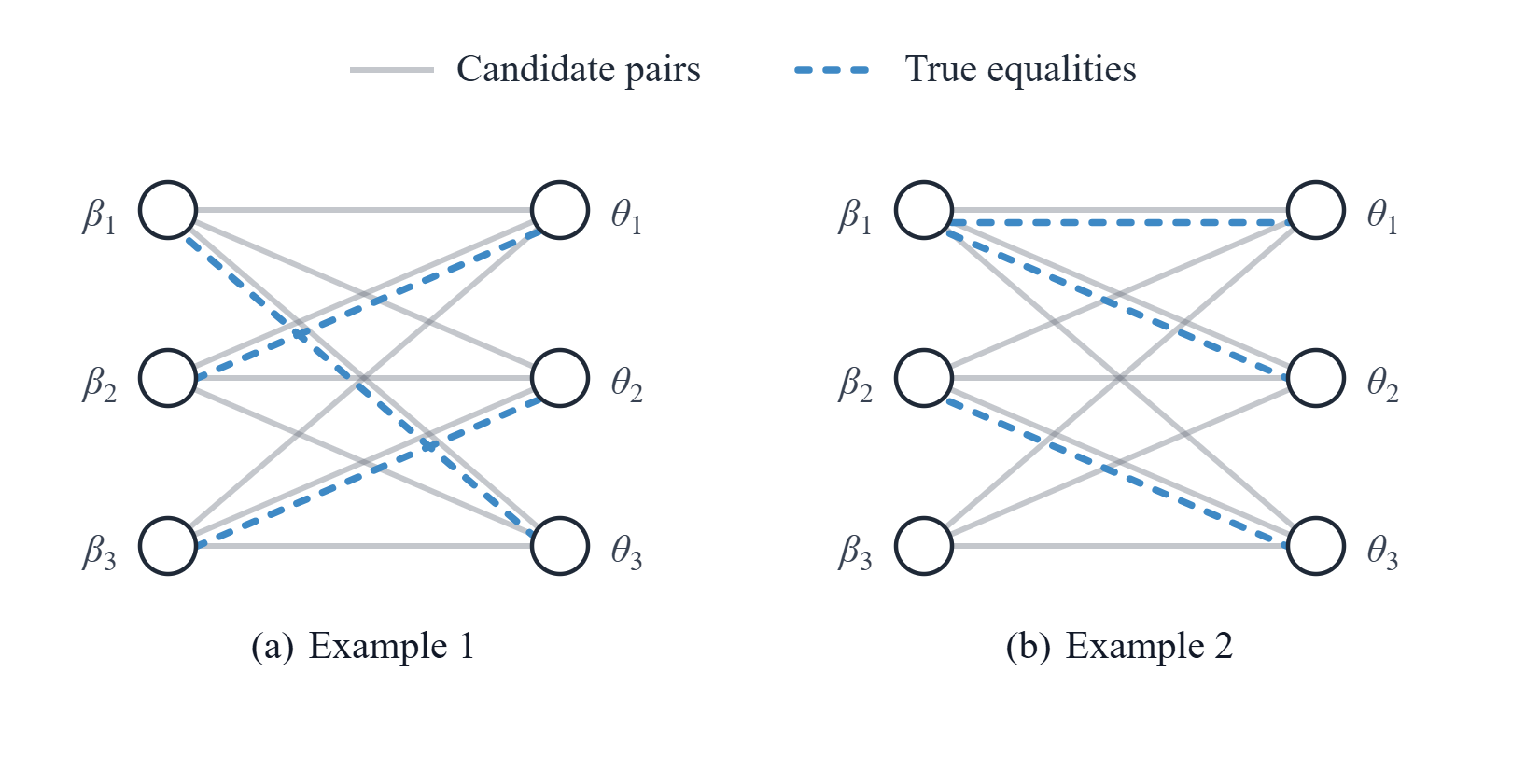} 
	\vspace{-1.0cm}   
	\caption{Bipartite-graph illustration of cross-semantic target--source correspondences in (a) Example 1 and (b) Example 2. Solid lines show all candidate $\beta_j-\theta_l$ fusion pairs, and dashed lines mark the true equalities in $\mathcal B$.}  
	\label{fig:ex1-mgraph}
\end{figure}

\noindent\textbf{Example 1.}
We set
$
\boldsymbol\beta^*=(1,2,3)^\top$
and $
\boldsymbol\theta^*=(2,3,1)^\top.
$ 
Here, the target and source share identical signal magnitudes but with permuted indices. The oracle equality set is given by
\[
\mathcal B=\{(j,\ell)\in[3]\times[3]:\beta_j^*=\theta_\ell^*\}
=\{(1,3),(2,1),(3,2)\},
\]
which corresponds to the constraints $\beta_1=\theta_3$, $\beta_2=\theta_1$, and $\beta_3=\theta_2$; see Figure~\ref{fig:ex1-mgraph}(a).

Formally, the oracle estimator solves the following constrained optimization problem:
\begin{equation}\label{eq:oracle-ex1}
	\min_{\boldsymbol{\beta},\boldsymbol{\theta}}
	\Bigl\{
	n_t^{-1}\|\mathbf Y^{(t)}-\mathbf X^{(t)}\boldsymbol\beta\|_2^2
	+
	n_s^{-1}\|\mathbf Y^{(s)}-\mathbf X^{(s)}\boldsymbol\theta\|_2^2
	\Bigr\}
	\quad\text{s.t.}\quad
	\beta_1=\theta_3,\ \beta_2=\theta_1,\ \beta_3=\theta_2.
\end{equation}
Let $\boldsymbol{\theta} = \mathbf{P}\boldsymbol{\beta}$, where the permutation matrix is defined as
\[
\mathbf P=
\begin{pmatrix}
	0 & 1 & 0\\
	0 & 0 & 1\\
	1 & 0 & 0
\end{pmatrix},
\]
yielding $\boldsymbol{\theta} = (\beta_2, \beta_3, \beta_1)^\top$. This transformation effectively re-aligns the source features to match the target coefficients. Consequently, \eqref{eq:oracle-ex1} reduces to an ordinary least-squares fit for $\boldsymbol\beta$ on the pooled data:
\[
\min_{\boldsymbol{\beta}}
\left\|
\begin{pmatrix}
	n_t^{-1/2}\mathbf Y^{(t)}\\
	n_s^{-1/2}\mathbf Y^{(s)}
\end{pmatrix}
-
\begin{pmatrix}
	n_t^{-1/2}\mathbf X^{(t)}\\
	n_s^{-1/2}\mathbf X^{(s)}\mathbf P
\end{pmatrix}
\boldsymbol{\beta}
\right\|_2^2.
\]

\begin{figure}[htbp]  
	\centering  
	\includegraphics[width=1.0\textwidth]{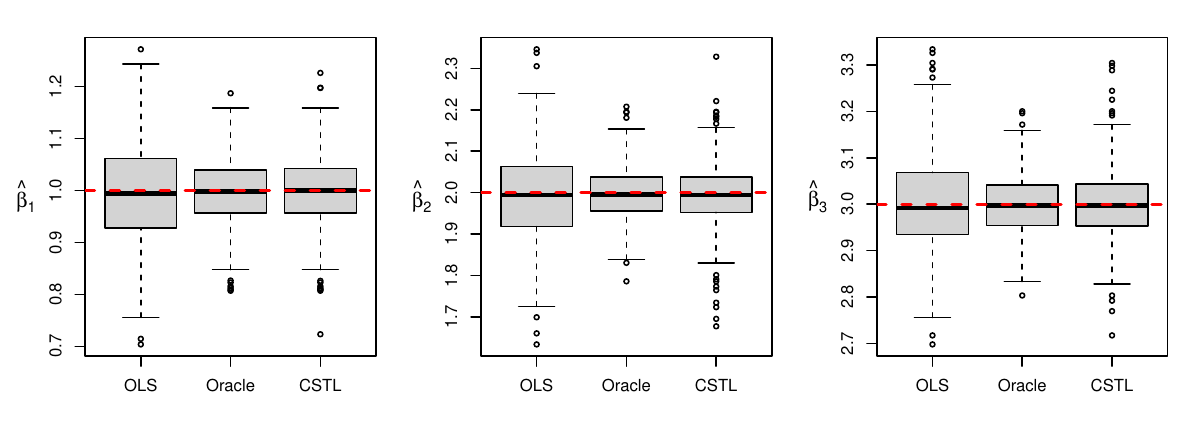}  
	\caption{
		Estimation performance in Example~1.
		Each panel shows boxplots of $\hat\beta_j$ ($j=1,2,3$) over 500 repetitions for OLS, the oracle estimator, and the data-driven CSTL.
		Horizontal dashed lines mark the ground-truth coefficients $\beta_j^*$.
	}
	\label{fig:ex1-boxplot}
\end{figure}

\begin{figure}[htbp]  
	\centering  
	\includegraphics[width=.9\textwidth]{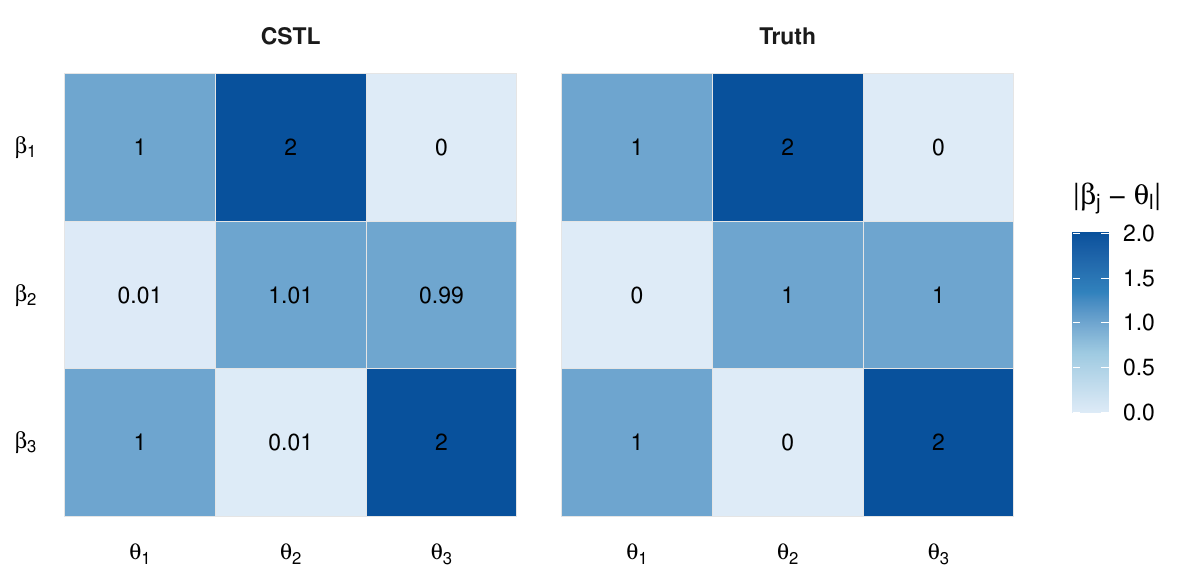}  
	\caption{Target--source coefficient differences in Example~1. Heatmaps of the empirical mean pairwise absolute differences   
		$|\hat\beta_j-\hat\theta_\ell|$ from CSTL over 500 repetitions (left)   
		and the true differences $|\beta_j^*-\theta_\ell^*|$ (right). Cell values report the corresponding absolute differences.}  
	\label{fig:ex1-heatmap}
\end{figure}

We compare three estimators of $\boldsymbol\beta$: the target-only OLS, the oracle, and the data-driven CSTL. Figure~\ref{fig:ex1-boxplot} demonstrates that the oracle produces noticeably narrower boxplots for $\hat\beta_1,\hat\beta_2,\hat\beta_3$ compared to OLS, reflecting the information gain obtained by borrowing strength from the source domain. Notably, the CSTL boxplots are
very close to those of the oracle, suggesting that the data-driven weights effectively recover the underlying transferable structure. This finding is further supported by Figure~\ref{fig:ex1-heatmap}, which compares the CSTL difference matrix, with entries $|\hat\beta_j-\hat\theta_\ell|$, against the ground truth $|\beta_j^*-\theta_\ell^*|$. The true matrix is zero solely at the indices in $\mathcal B=\{(1,3),(2,1),(3,2)\}$. The CSTL estimator reproduces this sparsity pattern, yielding near-zero values at these positions while maintaining non-zero values elsewhere.

We next consider a second example where the transferable structure is no longer one-to-one, meaning the oracle equalities cannot be represented by a single permutation matrix.

\noindent\textbf{Example 2.}
We set
$
\boldsymbol\beta^*=(1,2,3)^\top$
and
$
\boldsymbol\theta^*=(1,1,2)^\top.
$ 
In this scenario, the oracle equality set is given by
\[
\mathcal B=\{(j,\ell)\in[3]\times[3]:\beta_j^*=\theta_\ell^*\}
=\{(1,1), (1,2), (2,3)\}.
\]
This corresponds to the constraints $\beta_1=\theta_1$, $\beta_1=\theta_2$, and $\beta_2=\theta_3$,
as illustrated in Figure~\ref{fig:ex1-mgraph}(b). Formally, based on these constraints, the oracle estimator is defined as the solution to:
\begin{equation}\label{eq:oracle-ex2}	\min_{\boldsymbol{\beta},\boldsymbol{\theta}}	\Bigl\{	n_t^{-1}\|\mathbf Y^{(t)}-\mathbf X^{(t)}\boldsymbol\beta\|_2^2	+	n_s^{-1}\|\mathbf Y^{(s)}-\mathbf X^{(s)}\boldsymbol\theta\|_2^2	\Bigr\}	\quad\text{s.t.}\quad	\beta_1=\theta_1=\theta_2,\ \beta_2=\theta_3.
\end{equation}

Equivalently, \eqref{eq:oracle-ex2} can be written as
\begin{equation}\label{eq:oracle-ex2-rewrite}
	\min_{\boldsymbol{\beta}\in\mathbb R^3}
	\Bigl\{
	n_t^{-1}\|\mathbf Y^{(t)}-\mathbf X^{(t)}\boldsymbol\beta\|_2^2
	+
	n_s^{-1}\|\mathbf Y^{(s)}-(\mathbf X^{(s)}_{1}+\mathbf X^{(s)}_{2})\beta_1-\mathbf X^{(s)}_{3}\beta_2\|_2^2
	\Bigr\}.
\end{equation}

Intuitively, this formulation shows that $\beta_1$ is shared across two source features and the first target feature, and $\beta_2$ is shared across one source feature and the second target feature. In contrast, $\beta_3$ is unique to the target domain.

\begin{figure}[htbp]  
	\centering  
	\includegraphics[width=1.0\textwidth]{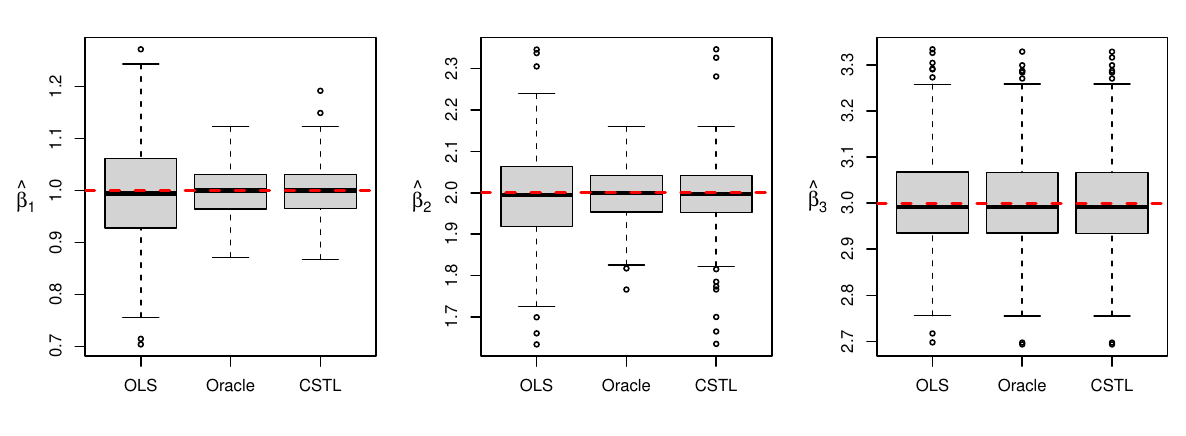}  
	\vspace{-0.9cm} 
	\caption{
		Estimation performance in Example~2.
		Each panel shows boxplots of $\hat\beta_j$ ($j=1,2,3$) over 500 repetitions for OLS, the oracle estimator, and the data-driven CSTL.
		Horizontal dashed lines mark the ground-truth coefficients $\beta_j^*$.
	}
	\label{fig:ex2-boxplot}
\end{figure}

\begin{figure}[htbp]  
	\centering  
	\includegraphics[width=.9\textwidth]{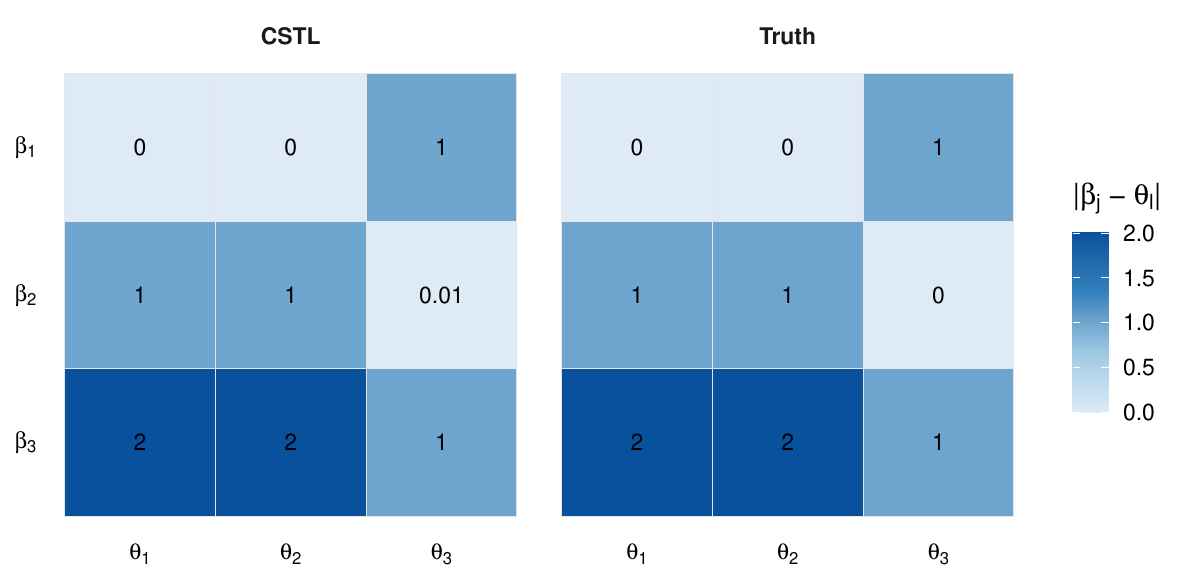}  
	\caption{Target--source coefficient differences in Example~2. Heatmaps of the empirical mean pairwise absolute differences   
		$|\hat\beta_j-\hat\theta_\ell|$ from CSTL over 500 repetitions (left)   
		and the true differences $|\beta_j^*-\theta_\ell^*|$ (right). Cell values report the corresponding absolute differences.}  
	\label{fig:ex2-heatmap}
\end{figure}

Similar to Example 1, Figure~\ref{fig:ex2-boxplot} compares the performance of the three methods. For $\beta_1$ and $\beta_2$, by integrating source information,
the oracle and CSTL estimators show reduced variance compared to OLS. For the non-transferable coefficient $\beta_3$, all three estimators behave similarly, indicating that CSTL correctly avoids negative transfer. Figure~\ref{fig:ex2-heatmap} further confirms that CSTL identifies the true equality set
$\mathcal B=\{(1,1),(1,2),(2,3)\}$ by shrinking the corresponding pairwise differences
to zero.

\section{Theoretical results}
\subsection{Conditions}
To establish the theoretical guarantees for the proposed CSTL, we begin by introducing three standard assumptions.

\begin{condition}[Design matrix]\label{cond1}
	For each $k\in\{t,s\}$, the rows of $\mathbf{X}^{(k)}$ are independent and identically distributed (i.i.d)  sub-Gaussian vectors with mean zero and covariance  matrix $\boldsymbol{\Sigma}^{(k)}$ satisfying  
	\begin{equation*}
		1 / c_0 \leq \lambda_{\min }\left(\boldsymbol{\Sigma}^{(k)}\right) \leq  \lambda_{\max }\left(\boldsymbol{\Sigma}^{(k)}\right) \leq c_0,
	\end{equation*}
	for some constant $c_0>0$.
\end{condition}

\begin{condition}[Noise variable]\label{cond2}
	For each  $k\in\{t,s\}$, the noises $\boldsymbol{\epsilon}_i^{(k)}$ ($1 \leq i \leq n_k$) are i.i.d 
	Gaussian random variables with mean zero and variance $\sigma_k^2$ satisfying $\max_{k \in \{t,s\}} \sigma_k^2 \leq c_\epsilon$ for some constant $c_\epsilon>0$. Furthermore, the noise vector $\boldsymbol{\epsilon}^{(k)}$ is independent of the design matrix $\mathbf{X}^{(k)}$.
\end{condition}

\begin{condition}[Minimum signal strength]\label{cond3}
	The minimal signal strengths for the target active coefficients and the non-transferable coefficient differences are bounded below as follows:
	\begin{equation*}
		\min_{j \in \mathcal{A}_t} |\beta_j^{*}| \geq \frac{3a}{2} \lambda_0 \quad \text{and} \quad 
		\min_{(j,l) \in \mathcal{B}^c} |\beta_{j}^{*} - \theta_{l}^{*}| \geq \frac{3a}{2} \lambda_1,
	\end{equation*}
	where $a>2$ is the SCAD penalty parameter and $\lambda_0,\lambda_1>0$ are the  regularization  parameters in CSTL.  
\end{condition}

Conditions \ref{cond1}–\ref{cond2} are standard in high-dimensional regression and transfer learning literature \citep{he2024transfusion,he2024adatrans}. The sub-Gaussian design assumption in Condition \ref{cond1} relaxes the Gaussian requirement, accommodating a broad class of light-tailed covariate distributions while allowing the target and source domains to have different feature distributions. Condition \ref{cond3} specifies the conventional minimal signal strength requirement, which is necessary to guarantee consistent support recovery in high-dimensional sparse regression \citep[e.g.,][]{zhao2006model, zhang2010nearly, he2024adatrans}.
\subsection{Main Theorems}
We now establish theoretical guarantees that connect the proposed CSTL method to the oracle solution in \eqref{oral}.  

\begin{theorem}[Ideal weights]\label{the1}
	Suppose Conditions \ref{cond1}-\ref{cond2} hold, and the sample sizes satisfy $n_t \gtrsim \log d_t$ and $n_s \gtrsim \log d_s$. Consider solving problem \eqref{original} with ideal weights $w_j = \mathbb{I}(j \in \mathcal{A}_t^c)$ and $w_{j,l} = \mathbb{I}((j,l) \in \mathcal{B})$, and 
	regularization parameters satisfying 
	\begin{equation*}
		\lambda_0 \geq C_1 \sqrt{ \frac{ \log d_t }{ n_t } } + \widetilde C_2 \cdot \frac{d_s}{d_t} \sqrt{ \frac{ \log d_s }{ n_s } }, \quad \lambda_1 \geq \max\left\{C_1' \sqrt{ \frac{ \log d_t }{ n_t } }, C_2' \sqrt{ \frac{ \log d_s }{ n_s } }\right\},
	\end{equation*}
	for some positive constants $C_1,\widetilde C_2,C_1',C_2'$. Then, with probability at least  $1 - c_1 \left(d_t^{-c_2}+ d_s^{-c_2}\right)$, the solution to problem \eqref{original} coincides with the oracle estimator in \eqref{oral}, where $c_1,c_2>0$ are universal constants.    
\end{theorem}

\textbf{Remark 1.}
Theorem~\ref{the1} establishes that with appropriate regularization parameters, CSTL equipped with the ideal weights recovers the oracle estimator with overwhelming probability.

\begin{theorem}[Data-driven weights]\label{the2}
	Suppose Conditions \ref{cond1}-\ref{cond3} hold, and the sample sizes satisfy 
	$n_t \gtrsim \log d_t$ and $n_s \gtrsim \log d_s$. 
	If there exist initial estimators $\hat{\boldsymbol{\beta}}_{\text{init}}, \hat{\boldsymbol{\theta}}_{\text{init}}$ and positive constants $C_1$, $\tilde{C}_2$, $C_1'$, $C_2'$ such that
	the regularization parameters  \( \lambda_0 \) and \( \lambda_1 \) satisfy 
	\begin{equation} 
		\left\|\hat{\boldsymbol{\beta}}_{\text{init}} - \boldsymbol{\beta}^{*} \right\|_{\infty} \vee \frac{1}{2} \Bigl(C_1 \sqrt{ \frac{ \log d_t }{ n_t } } 
		+ \widetilde C_2 \cdot \frac{d_s}{d_t} \sqrt{ \frac{ \log d_s }{ n_s } } \Bigr) \leq \frac{\lambda_0}{2} ,\label{eq:init1}
	\end{equation}
	\begin{equation}
		\max_{(j,l)\in [d_t] \times [d_s]} \left| (\hat{\beta}_{\text{init},j} - \hat{\theta}_{\text{init},l}) - (\beta_{j}^{*} - \theta_{l}^{*}) \right| \vee \frac{1}{2} \max\Bigl\{C_1'\sqrt{\frac{\log d_t}{n_t}},C_2'\sqrt{\frac{\log d_s}{n_s}} \Bigr\} \leq \frac{ \lambda_1}{2}, \label{eq:init2}
	\end{equation}
	then, with probability at least $1 - c_1 \left(d_t^{-c_2}+ d_s^{-c_2}\right)$, the oracle estimator $\hat{\boldsymbol{\beta}}_{\text{ora}}$ is attained by Algorithm~\ref{alg:CSTL},
	where $c_1$ and $c_2$ are universal positive constants.
\end{theorem}

\textbf{Remark 2.}
Theorem~\ref{the2} demonstrates that accurate initial estimators, combined with sufficiently strong target signals and large non-transferable coefficient differences, ensure the data-driven weights closely approximate their ideal counterparts in Theorem~\ref{the1}. Under these conditions and with appropriate regularization parameters, CSTL achieves the oracle estimator with high probability. This result validates the effectiveness of the proposed penalty design.

\section{ADMM Algorithm}\label{admm}
In this section, we employ the alternating direction method of multipliers (ADMM) to solve problem~\eqref{original}. 
We introduce auxiliary variables $z_j=\beta_j$ and $\delta_{j,l}=\beta_j-\theta_l$ to separate the absolute value terms from the quadratic loss.
Define 
$$P_n(\boldsymbol{\eta},\boldsymbol{z},\boldsymbol{\delta})=  \frac{1}{n_t}\|\mathbf{Y}^{(t)}-\mathbf{X}^{(t)}\boldsymbol{\beta}\|_2^2
+ \frac{1}{n_s}\|\mathbf{Y}^{(s)}-\mathbf{X}^{(s)}\boldsymbol{\theta}\|_2^2
+\lambda_0\sum_{j=1}^{d_t} w_j |z_j|
+\lambda_1\sum_{j=1}^{d_t}\sum_{l=1}^{d_s} w_{j,l}|\delta_{j,l}| ,
$$
where $\boldsymbol\eta=(\boldsymbol\beta^\top,\boldsymbol\theta^\top)^\top\in\mathbb R^{d_t+d_s}$, $\boldsymbol z=(z_1,\ldots,z_{d_t})^\top\in\mathbb R^{d_t}$, and $\boldsymbol\delta\in\mathbb R^{d_t d_s}$ stacks $\{\delta_{j,l}\}_{j,l}$ in the lexicographic order, namely,
\[
\boldsymbol\delta
=
(\delta_{1,1},\ldots,\delta_{1,d_s},\ 
\delta_{2,1},\ldots,\delta_{2,d_s},\ 
\ldots,\ 
\delta_{d_t,1},\ldots,\delta_{d_t,d_s})^\top .
\]
This leads to the following constrained problem:
\begin{equation}\label{eq:constrained}
	\begin{gathered} 
		\min_{\boldsymbol\eta,\boldsymbol z,\boldsymbol\delta}\quad
		P_n(\boldsymbol{\eta},\boldsymbol{z},\boldsymbol{\delta})\quad
		\text{s.t.}\quad
		\left\{ 
		\begin{aligned}
			& \beta_j - z_j = 0, \quad j=1,\ldots,d_t,\\
			& \beta_j - \theta_l - \delta_{j,l} = 0,\quad j=1,\ldots,d_t, \quad l=1,\ldots,d_s.
		\end{aligned}
		\right.
	\end{gathered}
\end{equation}

To express the constraints in \eqref{eq:constrained} in matrix form, define
$
\mathbf A=[\mathbf I_{d_t}\ \mathbf 0]\in\mathbb R^{d_t\times(d_t+d_s)},
$
so that $\mathbf A\boldsymbol\eta=\boldsymbol\beta$. Under the stacking order of $\boldsymbol\delta$, define $\mathbf D\in\mathbb R^{d_t d_s\times(d_t+d_s)}$ as
\[
\mathbf D=
\left(
\begin{array}{ccccc}
	\mathbf{1}_{d_s} & \mathbf{0} & \cdots & \mathbf{0} & -\mathbf{I}_{d_s} \\
	\mathbf{0} & \mathbf{1}_{d_s} & \cdots & \mathbf{0} & -\mathbf{I}_{d_s} \\
	\vdots & \vdots & \ddots & \vdots & \vdots \\
	\mathbf{0} & \mathbf{0} & \cdots & \mathbf{1}_{d_s} & -\mathbf{I}_{d_s}
\end{array}
\right),
\]
whose rows correspond to the pairwise differences $\beta_j-\theta_l$. Consequently,
\[
\mathbf D\boldsymbol\eta
=
(\beta_1-\theta_1,\ldots,\beta_1-\theta_{d_s},\ldots,
\beta_{d_t}-\theta_1,\ldots,\beta_{d_t}-\theta_{d_s})^\top.
\]
With these definitions, the constraints become $\boldsymbol z=\mathbf A\boldsymbol\eta$ and $\boldsymbol\delta=\mathbf D\boldsymbol\eta$.

Let $u_j$ and $v_{j,l}$ be the Lagrange multipliers associated with the constraints $\beta_j-z_j=0$ and $\beta_j-\theta_l-\delta_{j,l}=0$, respectively. 
Given $\rho_0,\rho_1>0$, the augmented Lagrangian function associated with \eqref{eq:constrained} is
\begin{align}\label{eq:augLag}
	\mathcal{Q}(\boldsymbol{\eta},\boldsymbol{z},\boldsymbol{\delta},\boldsymbol{u},\boldsymbol{v})
	=&P_n(\boldsymbol{\eta},\boldsymbol{z},\boldsymbol{\delta})
	+\sum_{j=1}^{d_t} u_j(\beta_j - z_j)
	+\sum_{j=1}^{d_t}\sum_{l=1}^{d_s} v_{j,l}(\beta_j - \theta_l - \delta_{j,l}) \nonumber\\
	&\ +\frac{\rho_0}{2}\sum_{j=1}^{d_t} (\beta_j - z_j)^2
	+\frac{\rho_1}{2}\sum_{j=1}^{d_t}\sum_{l=1}^{d_s} (\beta_j - \theta_l - \delta_{j,l})^2,
\end{align}
where
$\boldsymbol u=(u_1,\ldots,u_{d_t})^\top\in\mathbb R^{d_t}$ and $\boldsymbol v\in\mathbb R^{d_t d_s}$ stacks $\{v_{j,l}\}_{j,l}$ in the same order as $\boldsymbol\delta$.

Based on the augmented Lagrangian \eqref{eq:augLag}, the ADMM updates are given as follows. Given the current iterate $\bigl(\boldsymbol\eta^{(m)},\boldsymbol z^{(m)},\boldsymbol\delta^{(m)}, \boldsymbol u^{(m)},\boldsymbol v^{(m)}\bigr)$, the $(m+1)$-th iterate is computed by
\begin{align}
	\boldsymbol\eta^{(m+1)}
	&=\arg\min_{\boldsymbol\eta}\ 
	\mathcal Q\!\left(
	\boldsymbol\eta,\,
	\boldsymbol z^{(m)},\,
	\boldsymbol\delta^{(m)},\,
	\boldsymbol u^{(m)},\,
	\boldsymbol v^{(m)}
	\right), \label{admm:update:eta}\\
	\boldsymbol z^{(m+1)}
	&=\arg\min_{\boldsymbol z}\ 
	\mathcal Q\!\left(
	\boldsymbol\eta^{(m+1)},\,
	\boldsymbol z,\,
	\boldsymbol\delta^{(m)},\,
	\boldsymbol u^{(m)},\,
	\boldsymbol v^{(m)}
	\right), \label{admm:update:z}\\
	\boldsymbol\delta^{(m+1)}
	&=\arg\min_{\boldsymbol\delta}\ 
	\mathcal Q\!\left(
	\boldsymbol\eta^{(m+1)},\,
	\boldsymbol z^{(m+1)},\,
	\boldsymbol\delta,\,
	\boldsymbol u^{(m)},\,
	\boldsymbol v^{(m)}
	\right), \label{admm:update:delta}\\
	\boldsymbol u^{(m+1)}
	&=\boldsymbol u^{(m)}+\rho_0\bigl(\boldsymbol\beta^{(m+1)}-\boldsymbol z^{(m+1)}\bigr),
	\label{admm:update:u}\\
	\boldsymbol v^{(m+1)}
	&=\boldsymbol v^{(m)}+\rho_1\bigl(\mathbf D\boldsymbol\eta^{(m+1)}-\boldsymbol\delta^{(m+1)}\bigr),
	\label{admm:update:v}
\end{align}
where $\boldsymbol\eta^{(m+1)}=(\boldsymbol\beta^{(m+1)\top},\boldsymbol\theta^{(m+1)\top})^\top$. Each subproblem in \eqref{admm:update:eta}--\eqref{admm:update:delta} admits a closed-form solution, which is summarized below.

We begin with the $\boldsymbol\eta$-update in \eqref{admm:update:eta}. Keeping $(\boldsymbol z^{(m)},\boldsymbol\delta^{(m)},\boldsymbol u^{(m)},\boldsymbol v^{(m)})$ fixed and removing terms that do not depend on $\boldsymbol\eta$, the subproblem becomes
\begin{align}\label{eq:eta-subprob-scalar}
	\boldsymbol\eta^{(m+1)}
	=\arg\min_{\boldsymbol\eta}\ 
	&\frac{1}{n_t}\bigl\|\mathbf Y^{(t)}-\mathbf X^{(t)}\boldsymbol\beta\bigr\|_2^2
	+\frac{1}{n_s}\bigl\|\mathbf Y^{(s)}-\mathbf X^{(s)}\boldsymbol\theta\bigr\|_2^2 
	+\frac{\rho_0}{2}\sum_{j=1}^{d_t}\Bigl(\beta_j-z_j^{(m)}+\rho_0^{-1}u_j^{(m)}\Bigr)^2\nonumber\\
	&+\frac{\rho_1}{2}\sum_{j=1}^{d_t}\sum_{l=1}^{d_s}
	\Bigl(\beta_j-\theta_l-\delta_{j,l}^{(m)}+\rho_1^{-1}v_{j,l}^{(m)}\Bigr)^2 .
\end{align}
Using $\boldsymbol\beta=\mathbf A\boldsymbol\eta$ and the definition of $\mathbf D$, we can rewrite \eqref{eq:eta-subprob-scalar} as
\begin{eqnarray}\label{eq:eta-subprob}
	\boldsymbol\eta^{(m+1)}
	=\arg\min_{\boldsymbol\eta}\ 
	\|\mathbf Y-\mathbf X\boldsymbol\eta\|_2^2
	&+\frac{\rho_0}{2}\Bigl\|\mathbf A\boldsymbol\eta-\boldsymbol z^{(m)}+\rho_0^{-1}\boldsymbol u^{(m)}\Bigr\|_2^2\nonumber\\
	&+\frac{\rho_1}{2}\Bigl\|\mathbf D\boldsymbol\eta-\boldsymbol\delta^{(m)}+\rho_1^{-1}\boldsymbol v^{(m)}\Bigr\|_2^2,
\end{eqnarray}
where $\mathbf X=\mathrm{diag}\!\left(\mathbf X^{(t)}/\sqrt{n_t},\ \mathbf X^{(s)}/\sqrt{n_s}\right)$,
and $\mathbf Y$ is defined in \eqref{eq:pooled-design}. This is a strictly convex quadratic problem. Taking the first-order optimality condition of \eqref{eq:eta-subprob} gives
\begin{eqnarray}\label{eq:eta-closed}
	\boldsymbol\eta^{(m+1)}
	&=&
	\Bigl(
	2\mathbf X^\top\mathbf X
	+\rho_0\mathbf A^\top\mathbf A
	+\rho_1\mathbf D^\top\mathbf D
	\Bigr)^{-1}\nonumber\\
	&&\times
	\Bigl\{
	2\mathbf X^\top\mathbf Y
	+\mathbf A^\top(\rho_0\boldsymbol z^{(m)}-\boldsymbol u^{(m)})
	+\mathbf D^\top(\rho_1\boldsymbol\delta^{(m)}-\boldsymbol v^{(m)})
	\Bigr\}.
\end{eqnarray}

Next, we update $\boldsymbol z$ via \eqref{admm:update:z}. The problem is separable across coordinates. For each $j=1,\ldots,d_t$, we have
\[
z_j^{(m+1)}
=\arg\min_{z\in\mathbb R}\ 
\lambda_0 w_j|z|
+\frac{\rho_0}{2}\Bigl(z-\beta_j^{(m+1)}-\rho_0^{-1}u_j^{(m)}\Bigr)^2,
\]
and thus
\begin{equation}\label{eq:z-closed}
	z_j^{(m+1)}
	=\mathsf S\!\left(\beta_j^{(m+1)}+\rho_0^{-1}u_j^{(m)},\
	\frac{\lambda_0 w_j}{\rho_0}\right),
\end{equation}
where $\mathsf S(x,\tau)=\mathrm{sign}(x)\max\{|x|-\tau,0\}$.

Finally, we update $\boldsymbol\delta$ via \eqref{admm:update:delta}. This subproblem is separable over pairs $(j,l)$. For each $j=1,\ldots,d_t$ and $l=1,\ldots,d_s$, we obtain that
\[
\delta_{j,l}^{(m+1)}
=\arg\min_{\delta\in\mathbb R}\ 
\lambda_1 w_{j,l}|\delta|
+\frac{\rho_1}{2}\Bigl(\delta-(\beta_j^{(m+1)}-\theta_l^{(m+1)})-\rho_1^{-1}v_{j,l}^{(m)}\Bigr)^2,
\]
which yields
\begin{equation}\label{eq:delta-closed}
	\delta_{j,l}^{(m+1)}
	=
	\mathsf S\!\left(
	\beta_j^{(m+1)}-\theta_l^{(m+1)}+\rho_1^{-1}v_{j,l}^{(m)},\
	\frac{\lambda_1 w_{j,l}}{\rho_1}
	\right).
\end{equation}

We adopt the standard ADMM stopping criterion based on primal and dual residuals.
Define the primal residuals
\[
\boldsymbol r_0^{(m+1)}=\mathbf A\boldsymbol\eta^{(m+1)}-\boldsymbol z^{(m+1)},\qquad
\boldsymbol r_1^{(m+1)}=\mathbf D\boldsymbol\eta^{(m+1)}-\boldsymbol\delta^{(m+1)},
\]
and the dual residuals
\[
\boldsymbol s_0^{(m+1)}=\rho_0\,\mathbf A^\top\!\bigl(\boldsymbol z^{(m+1)}-\boldsymbol z^{(m)}\bigr),\qquad
\boldsymbol s_1^{(m+1)}=\rho_1\,\mathbf D^\top\!\bigl(\boldsymbol\delta^{(m+1)}-\boldsymbol\delta^{(m)}\bigr).
\]
Following the stopping rule in \citet{boyd2011distributed}, the ADMM iterations are terminated
when both primal and dual feasibility are sufficiently small, i.e.,
\begin{equation}\label{eq:stoprule}
	\max\{\|\boldsymbol r_0^{(m+1)}\|_2,\ \|\boldsymbol r_1^{(m+1)}\|_2\}
	\le \varepsilon_{\mathrm{pri}}
	\quad\text{and}\quad
	\max\{\|\boldsymbol s_0^{(m+1)}\|_2,\ \|\boldsymbol s_1^{(m+1)}\|_2\}
	\le \varepsilon_{\mathrm{dual}},
\end{equation}
where $\varepsilon_{\mathrm{pri}}$ and $\varepsilon_{\mathrm{dual}}$ are pre-specified tolerances.
In summary, our ADMM procedure for solving \eqref{original} is summarized in Algorithm~\ref{alg:admm_final}.

\begin{algorithm}[H]
	\setlength{\baselineskip}{20pt} 
	\caption{ADMM for solving \eqref{original}}
	\label{alg:admm_final}
	\textbf{Input:} Data $\{(\mathbf X^{(t)},\mathbf Y^{(t)}),(\mathbf X^{(s)},\mathbf Y^{(s)})\}$; $(\lambda_0,\lambda_1)$; $(\rho_0,\rho_1)$; $(\varepsilon_{\rm pri},\varepsilon_{\rm dual})$; $M_{\max}$.\\
	\textbf{Output:} $\boldsymbol\eta$. \\
	\textbf{Step 1.} Initialize
	$\boldsymbol\eta^{(0)}$,
	$\boldsymbol z^{(0)}$, $\boldsymbol\delta^{(0)}$,
	$\boldsymbol u^{(0)}$, and $\boldsymbol v^{(0)}$. \\
	\textbf{Step 2.} For $m=0,1,\ldots,M_{\max}-1$, repeat:\par\vspace{0.9em}
	\hspace*{2.5em}\parbox{0.9\linewidth}{%
		$\boldsymbol\eta^{(m+1)} \leftarrow$ update by \eqref{eq:eta-closed}.\\
		$\boldsymbol z^{(m+1)} \leftarrow$ update by \eqref{eq:z-closed}.\\
		$\boldsymbol\delta^{(m+1)} \leftarrow$ update by \eqref{eq:delta-closed}.\\
		$\boldsymbol u^{(m+1)} \leftarrow$ update by \eqref{admm:update:u}.\\
		$\boldsymbol v^{(m+1)} \leftarrow$ update by \eqref{admm:update:v}.\\
		\textbf{if} stopping rule \eqref{eq:stoprule} holds \textbf{then break}.%
	}
\end{algorithm}

\section{Simulations and Empirical Studies}\label{sec:simulation}
\subsection{Simulation Studies}
In this section, we conduct simulation studies to evaluate the performance of our proposed method CSTL (implemented via Algorithm~\ref{alg:CSTL}) and the oracle estimator (with closed form given in Proposition~\ref{pro-oracle-closed}), in comparison with three existing methods: TransLasso \citep{li2022transfer}, TransGLM \citep{tian2023transfer} and Lasso \citep{tibshirani1996regression}. For CSTL, the regularization parameters $(\lambda_0, \lambda_1)$ are selected by minimizing the Bayesian Information Criterion (BIC). Following \citet{tang2016fused}, the BIC for a given $(\lambda_0, \lambda_1)$ is defined as
\begin{eqnarray*}
	\mathrm{BIC}(\lambda_0, \lambda_1)
	&=& \tfrac{N}{2} \Big[
	\log \!\left(
	\tfrac{1}{n_t} \big\|\mathbf{Y}^{(t)} - \mathbf{X}^{(t)} \hat{\boldsymbol{\beta}}_{\mathrm{cst}} \big\|_2^2
	\right)
	+ \log \!\left(
	\tfrac{1}{n_s} \big\|\mathbf{Y}^{(s)} - \mathbf{X}^{(s)} \hat{\boldsymbol{\theta}}_{\mathrm{cst}} \big\|_2^2
	\right)
	\Big]\\
	&& + \mathrm{df}(\lambda_0, \lambda_1) \cdot \log(N),  
\end{eqnarray*}
where $N=n_t+n_s$, and the degrees of freedom $\mathrm{df}(\lambda_0, \lambda_1)$ are defined as the number of distinct coefficients in the estimated vector  $\hat{\boldsymbol{\eta}}=\left(\hat{\boldsymbol{\beta}}_{\text{cst}}^{\top},\hat{\boldsymbol{\theta}}_{\text{cst}}^{\top} \right)^{\top}$, so that coefficients fused to the same value are counted once. For competing methods, we adopt the tuning and implementation details as suggested in their original papers.

All methods are evaluated on the target domain using two metrics:
(1) the sum of squared estimation errors (SSE)
defined as 
$\|\hat{\boldsymbol{\beta}}-\boldsymbol{\beta}^{*}\|_2^2$, where \( \hat{\boldsymbol{\beta}} \) is the estimated target coefficient vector;
(2) the mean squared prediction errors (MSE), calculated as 
$\frac{1}{100}\sum_{i=1}^{100}(y_i-\mathbf{x}_i^\top\hat{\boldsymbol{\beta}})^2$
over $100$ independently generated test samples ${(\mathbf{x}_i, y_i)}_{i=1}^{100}$ drawn from the target model. Each simulation is repeated $100$ times, 
and the average results are presented in Figures~\ref{S1}--\ref{S3mse} and Table~\ref{S4}.

\subsubsection{Simulation Settings}
We generate samples from the following linear regression models:
\begin{equation*} \mathbf{Y}^{(t)}=\mathbf{X}^{(t)}\boldsymbol{\beta}^{*}+  \boldsymbol{\epsilon}^{(t)} \text{ and} \quad \mathbf{Y}^{(s)}=\mathbf{X}^{(s)} \boldsymbol{\theta}^{*}+  \boldsymbol{\epsilon}^{(s)}.
\end{equation*}
For $k \in \{t, s\}$, $\boldsymbol{\epsilon}^{(k)} \sim N\left(\mathbf{0},  \boldsymbol{I}_{n_k}\right)$, and each row of the covariate matrix \( \mathbf{X}^{(k)} \in \mathbb{R}^{n_k \times d_k} \) is independently generated from $N_{d_k}(\mathbf{0}, \boldsymbol{\Sigma}^{(k)})$, with $\Sigma^{(k)}_{j_1, j_2}=0.5^{\left|j_2-j_1\right|}$, for $j_1, j_2=1, \ldots, d_k$. 
The source sample size was fixed at $n_s=500$, 
and the target sample size $n_t$  was varied among $ \{200,300,400\}$.

We design four simulation settings to evaluate different transfer learning scenarios. Settings 1–3 operate under a homogeneous feature space, ensuring the applicability of existing transfer methods that require semantic alignment. Within this subset, Settings 1 and 2 assess robustness to cross-semantic signal similarity, while Setting 3 investigates partial information sharing—incorporating a permutation variant as a stress test. Finally, Setting 4 evaluates robustness against differing target-source dimensional configurations.

\noindent\textbf{Setting 1.}
Let $d_t=d_s=600$ and define the target coefficient
\begin{equation*}
	\boldsymbol{\beta}^{*}=(\underbrace{1,\ldots,1}_{30},\underbrace{0,\ldots,0}_{d_t-30})^\top.
\end{equation*}
Then
$\mathcal{A}_t=\{1,\ldots,30\},\quad\mbox{\rm and}\quad
\mathcal{A}_t^c=\{1,\ldots,d_t\}\setminus \mathcal{A}_t.  $
For $m\in\{0,1,2,3,4\}$, draw disjoint subsets
$I_0\subset \mathcal{A}_t$ and $I_1\subset \mathcal{A}_t^{c}$ of size $m$ uniformly without replacement, and define source coefficient $\boldsymbol{\theta}^{*}$ by
\begin{equation*}
	\theta^{*}_j =
	\begin{cases}
		0, & j\in I_0,\\[2pt]
		1, & j\in I_1,\\[2pt]
		\beta_j^{*}, & \text{otherwise}.
	\end{cases}
\end{equation*}

\noindent\textbf{Setting 2.}
Let $d_t=d_s=600$ and define
\begin{equation*}
	\boldsymbol{\beta}^{*}=\bigl(\underbrace{-4,-3,-2,-1,1,2,3,4}_{8},\underbrace{0,\ldots,0}_{d_t-8}\bigr)^\top.
\end{equation*}
Then $\mathcal{A}_t=\{1,\ldots,8\}$
and
$\mathcal{A}_t^c=\{1,\ldots,d_t\}\setminus \mathcal{A}_t .$
Similarly, choose $m\in\{0,1,2,3,4\}$ and draw disjoint subsets $I_0\subset \mathcal{A}_t$ and $I_1\subset \mathcal{A}_t^{c}$ with $\left|I_0\right|=\left|I_1\right|=m$ uniformly without replacement.  
Construct source coefficients by
\begin{equation*}
	(\boldsymbol{\theta}^{*})_{I_0}=\mathbf{0},\qquad
	(\boldsymbol{\theta}^{*})_{I_1}=(\boldsymbol{\beta}^{*})_{I_0},\qquad
	(\boldsymbol{\theta}^{*})_{(I_0\cup I_1)^c}=(\boldsymbol{\beta}^{*})_{(I_0\cup I_1)^c}.
\end{equation*}

\noindent \textbf{Setting 3.} Let $d_t=d_s=600$. The target coefficient vector and the perturbation vector are defined as
\begin{equation*}
	\boldsymbol{\beta}^{*} = (\underbrace{1,\ldots,1}_{8},\underbrace{0,\ldots,0}_{d_t - 8})^\top,\quad
	\mbox{\rm and}\quad
	\boldsymbol{\delta}^{(s)} = (\underbrace{\delta_1^{(s)},\ldots,\delta_4^{(s)}}_{4}, \underbrace{0,\ldots,0}_{d_t - 4})^\top,
\end{equation*}
where $\delta_j^{(s)} \sim \mathcal{N}(h, (h/3)^2)$ for $j = 1,\ldots,4$,
and 
the strength of partial heterogeneity $h$ across tasks varies among 
$\{0.0,0.1,\ldots,0.5\}$. We consider two variants: (i) \textbf{No permutation with partial heterogeneity}: $\boldsymbol{\theta}^{*}=\boldsymbol{\beta}^{*}+\boldsymbol{\delta}^{(s)}$; (ii) \textbf{Global permutation with partial heterogeneity}: $\boldsymbol{\theta}^{*}=\mathbf{P}^{(s)}\bigl(\boldsymbol{\beta}^{*}+\boldsymbol{\delta}^{(s)}\bigr)$, where $\mathbf{P}^{(s)}$ is a $d_t\times d_t$ permutation matrix (one nonzero per row and column).

\noindent\textbf{Setting 4.} To simulate covariate dimensional mismatch, we fix the target dimension at $d_t=600$, while the source dimension $d_s$ varies among $\{550,600,650\}$. We set
\begin{equation*}
	\boldsymbol{\beta}^{*} = \bigl(\underbrace{1,\ldots,1}_{8},\underbrace{0,\ldots,0}_{d_t-8}\bigr)^\top,\quad
	\mbox{\rm and}\quad 
	\boldsymbol{\theta}^{*} = \bigl(\underbrace{1+\delta^{(s)}_1,\ldots,1+\delta^{(s)}_4}_{4},\underbrace{1,\ldots,1}_{4},\underbrace{0,\ldots,0}_{d_s-8}\bigr)^\top,
\end{equation*}
where \( \delta_j^{(s)} \sim \mathcal{N}(0.5, (0.5/3)^2) \) for \( j = 1, \ldots, 4 \). 

\subsubsection{Simulation Results}

\begin{figure}[!htbp]
	\centering
	\includegraphics[width=1\textwidth]{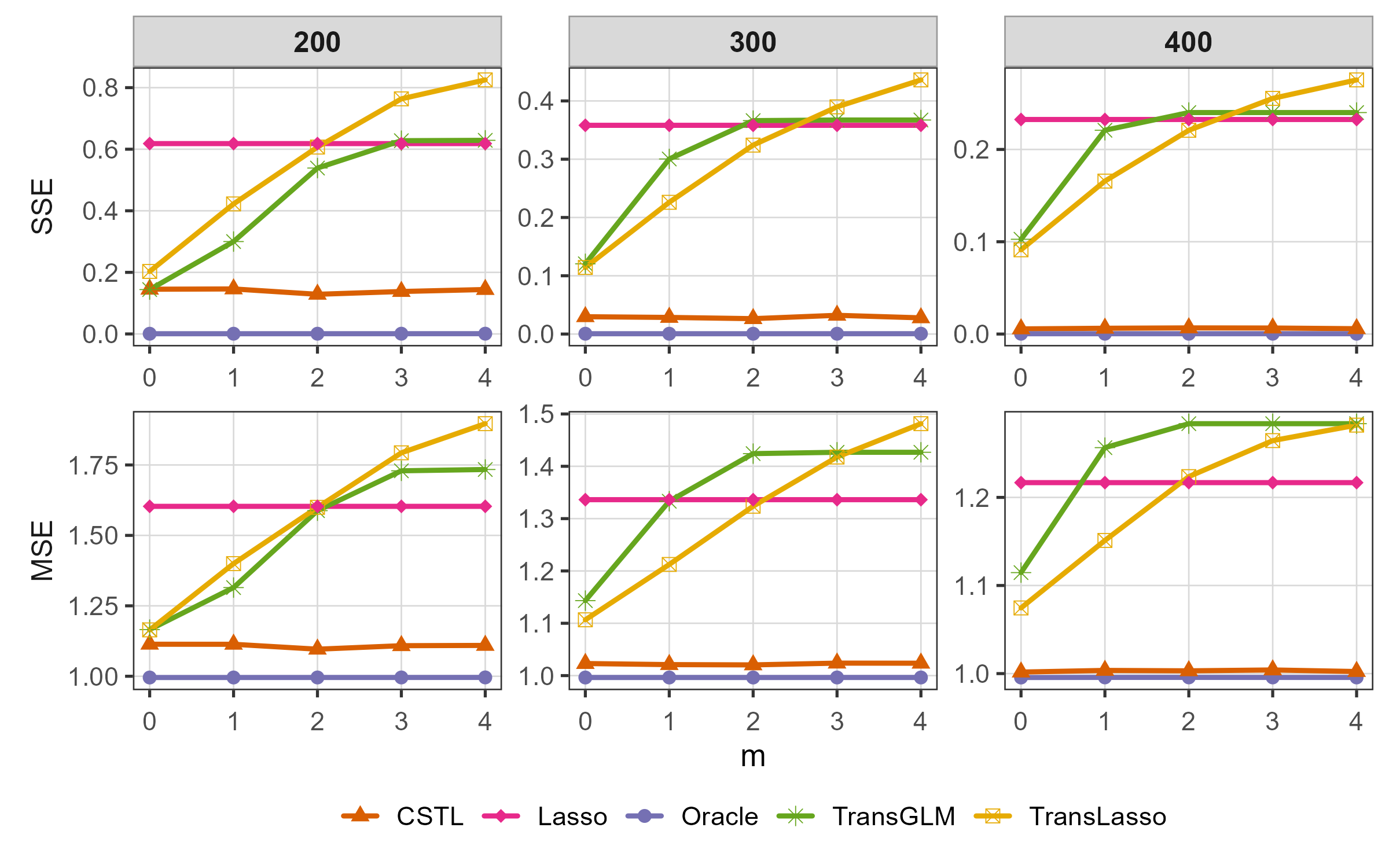}
	\caption{Average SSE and MSE versus $m$ under Setting 1.}
	\label{S1}
\end{figure}

\begin{figure}[!htbp]
	\centering
	\includegraphics[width=1\textwidth]{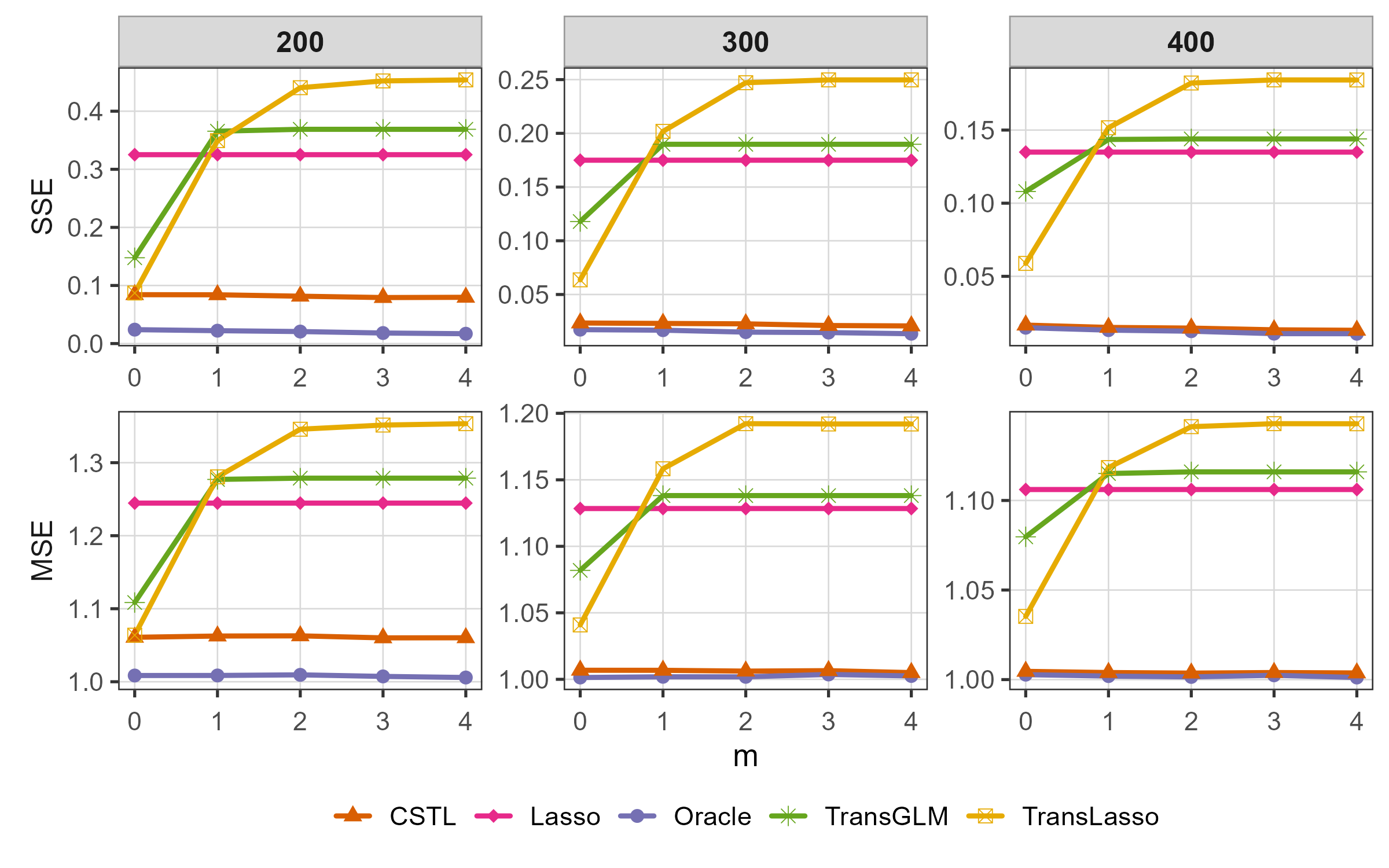}
	\caption{Average SSE and MSE versus $m$ under Setting 2.}
	\label{S2}
\end{figure}

Figures~\ref{S1}--\ref{S2} present the average SSE and MSE for Settings 1-2, with three main findings: 
(i) Robustness to Semantic Similarity ($m$):
For any fixed $n_t$, 
the performance of CSTL remains nearly invariant to changes in $m$ and consistently surpasses Lasso, demonstrating empirical robustness under varying cross-semantic similarity. In contrast, while TransLasso and TransGLM are competitive at $m=0$, their errors increase significantly -- often exceeding those of Lasso -- as $m$ grows, indicating a lack of robustness.
(ii) Benefit of Increased Target Data ($n_t$):
As $n_t$ increases, all methods improve. CSTL, in particular, steadily approaches the oracle performance. This aligns with the methodological rationale that a more accurate initial estimator yields more reliable transferability weights, enabling a sharper distinction between transferable and non-transferable components.
(iii) Consistency Between Metrics:
The observed MSE trends are in strong agreement with the SSE results, reinforcing the above conclusions.

\begin{figure}[!htbp]
	\centering
	\includegraphics[width=1\textwidth]{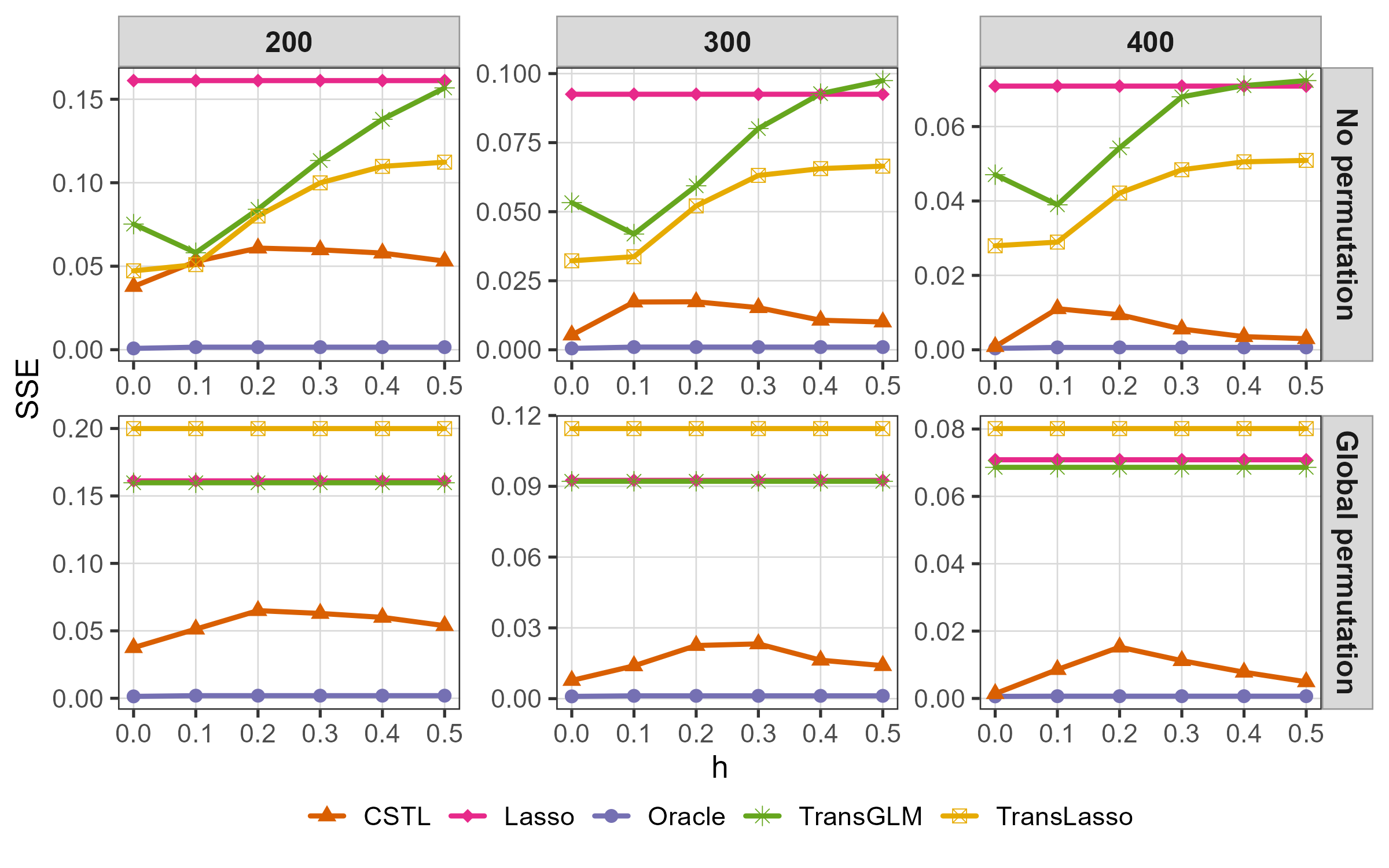}
	\caption{SSE versus heterogeneity level $h$ under Setting 3.}
	\label{S3sse}
\end{figure}

\begin{figure}[!htbp]
	\centering
	\includegraphics[width=1\textwidth]{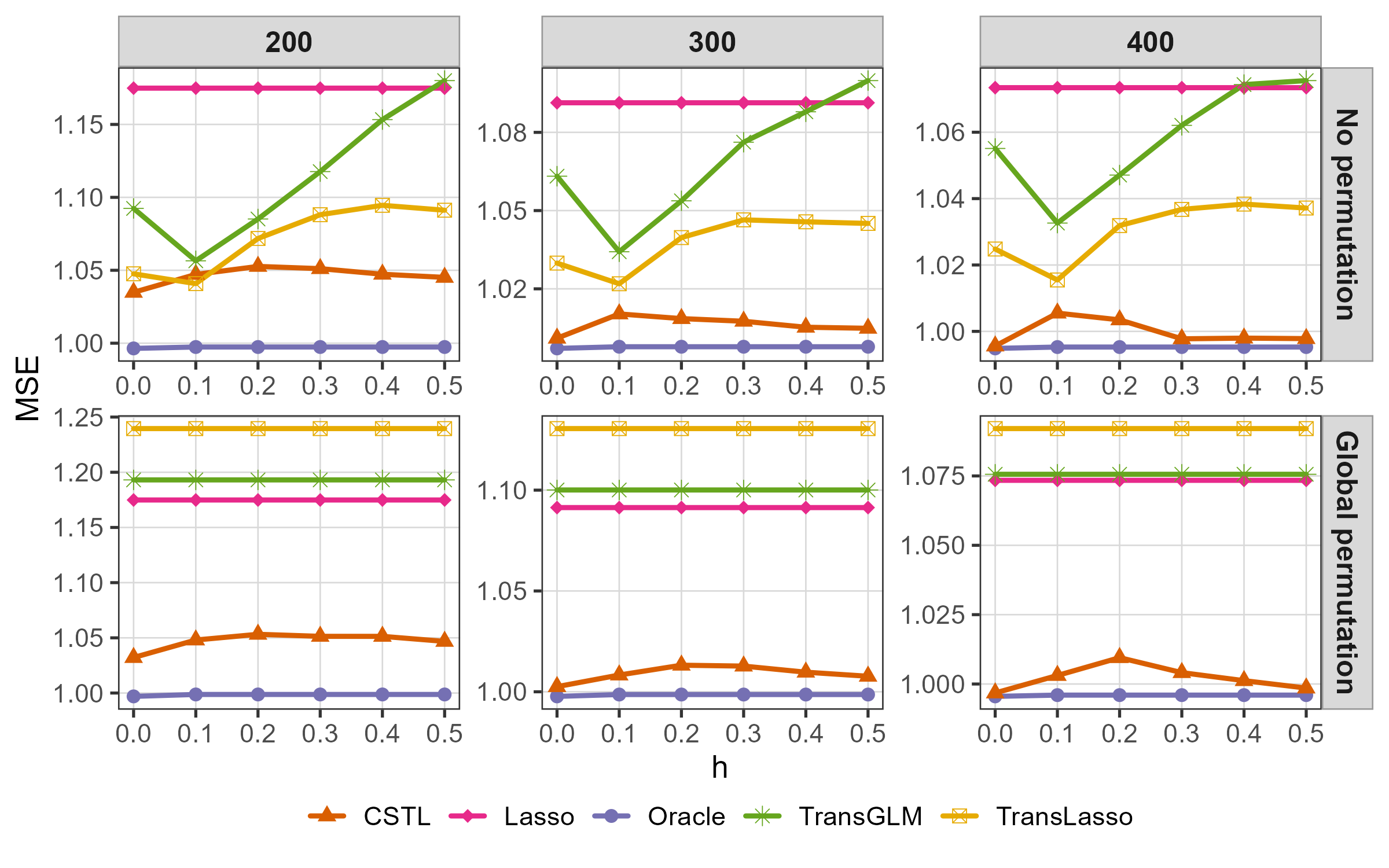}
	\caption{MSE versus heterogeneity level $h$ under Setting 3.}
	\label{S3mse}
\end{figure}

Figures~\ref{S3sse}--\ref{S3mse}  present the average SSE and MSE for Settings 3 across different heterogeneity levels \(h\). The key observations are as follows:
\begin{itemize}
	\item[(1)] \textbf{No permutation.} 
	CSTL consistently outperforms Lasso across all levels of heterogeneity $h$, exhibiting a characteristic pattern: errors initially increase slightly before declining as $h$ grows. This non-monotonic trend can be explained by the method's transferability assessment mechanism. Under moderate heterogeneity, a small proportion of non-transferable signals may be misclassified as transferable, leading to a slight performance dip. However, as heterogeneity increases further, the distinction between transferable and non-transferable components becomes sharper, allowing CSTL to more effectively filter out non-transferable information, thereby improving estimation accuracy.
	
	In contrast, TransLasso and TransGLM perform competitively only under mild heterogeneity ($h\le 0.1$). Once $h$ exceeds a moderate threshold (e.g., $h\ge 0.2$), their estimation errors increase significantly and remain elevated, indicating a lack of robustness to stronger heterogeneity.
	\item[(2)]  \textbf{Global permutation.}
	Across all levels of heterogeneity $h$ and sample sizes, CSTL performs universally superior to all benchmarks and remains competitive with the Oracle, exhibiting an error pattern over $h$ consistent with the non-permuted scenario. By comparison, TransLasso and TransGLM yield results comparable to or worse than Lasso, demonstrating that they are incapable of harnessing useful knowledge from the source under global permutation.
\end{itemize}

Table \ref{S4} presents the results for Setting 4, where the source and target feature dimensions differ ($d_s\neq d_t$). CSTL consistently outperforms Lasso across all dimensions ($d_s$) and target sample sizes ($n_t$).
Its SSE and MSE decrease as $n_t$ grows, 
closely approaching the oracle performance.
This result confirms CSTL's capability to reliably identify and transfer useful signals even under heterogeneous feature spaces. In contrast, TransLasso and TransGLM are rendered inapplicable in this setting due to their fundamental requirement of feature-space homogeneity ($d_s =d_t$).

\begin{table}[!h]
	\centering
	\renewcommand{\arraystretch}{0.6}
	\caption{Average estimation error (SSE) and prediction error (MSE) in Setting 4.}
	\label{S4}
	\begin{tabular}{ccccccc}
		\toprule
		$n_t$ & $d_s$ & Lasso & TransLasso & TransGLM& CSTL &Oracle\\\midrule
		\multicolumn{7}{c}{SSE}\\\hline
		~ & 550 & 0.16117 &--&--& 0.05155 & 0.00140  \\ 
		200 & 600 & 0.16117 & 0.11230 & 0.15679 & 0.05309 & 0.00152 \\ 
		~ & 650 & 0.16117 &--&--& 0.04995 & 0.00170  \\ \hline
		~ & 550 & 0.09252 &--&--& 0.00986 & 0.00097 \\ 
		300 & 600 & 0.09252 & 0.06649 & 0.09746 & 0.01007 & 0.00098 \\ 
		~ & 650 & 0.09252 &--&--& 0.00953 & 0.00093 \\ \hline
		~ & 550 & 0.07087 &--&--& 0.00419 & 0.00062 \\ 
		400 & 600 & 0.07087 & 0.05090 & 0.07235 & 0.00298 & 0.00064 \\ 
		~ & 650 & 0.07087 &--&--& 0.00272 & 0.00057 \\ \hline
		\multicolumn{7}{c}{MSE}\\\hline
		~ & 550 & 1.17488 &--&--& 1.04645 & 0.99677 \\ 
		200 & 600 & 1.17488 & 1.09118 & 1.18005 & 1.04512 & 0.99739 \\ 
		~ & 650 & 1.17488 &--&--& 1.04824 & 0.99798 \\ \hline
		~ & 550 & 1.09135 &--&--& 1.00374 & 0.99874  \\ 
		300 & 600 & 1.09135 & 1.04508 & 1.09985 & 1.00486 & 0.99779 \\ 
		~ & 650 & 1.09135 &--&--& 1.00386 & 0.99797 \\ \hline
		~ & 550 & 1.07342 &--&--& 0.99916 & 0.99624 \\ 
		400 & 600 & 1.07342 & 1.03716 & 1.07555 & 0.99782 & 0.99528 \\ 
		~ & 650 & 1.07342 &--&--& 0.99889 & 0.99574\\
		\bottomrule
	\end{tabular}
\end{table}

\subsection{Real Data}

\begin{figure}[!htbp]
	\centering
	\includegraphics[width=1.0\textwidth]{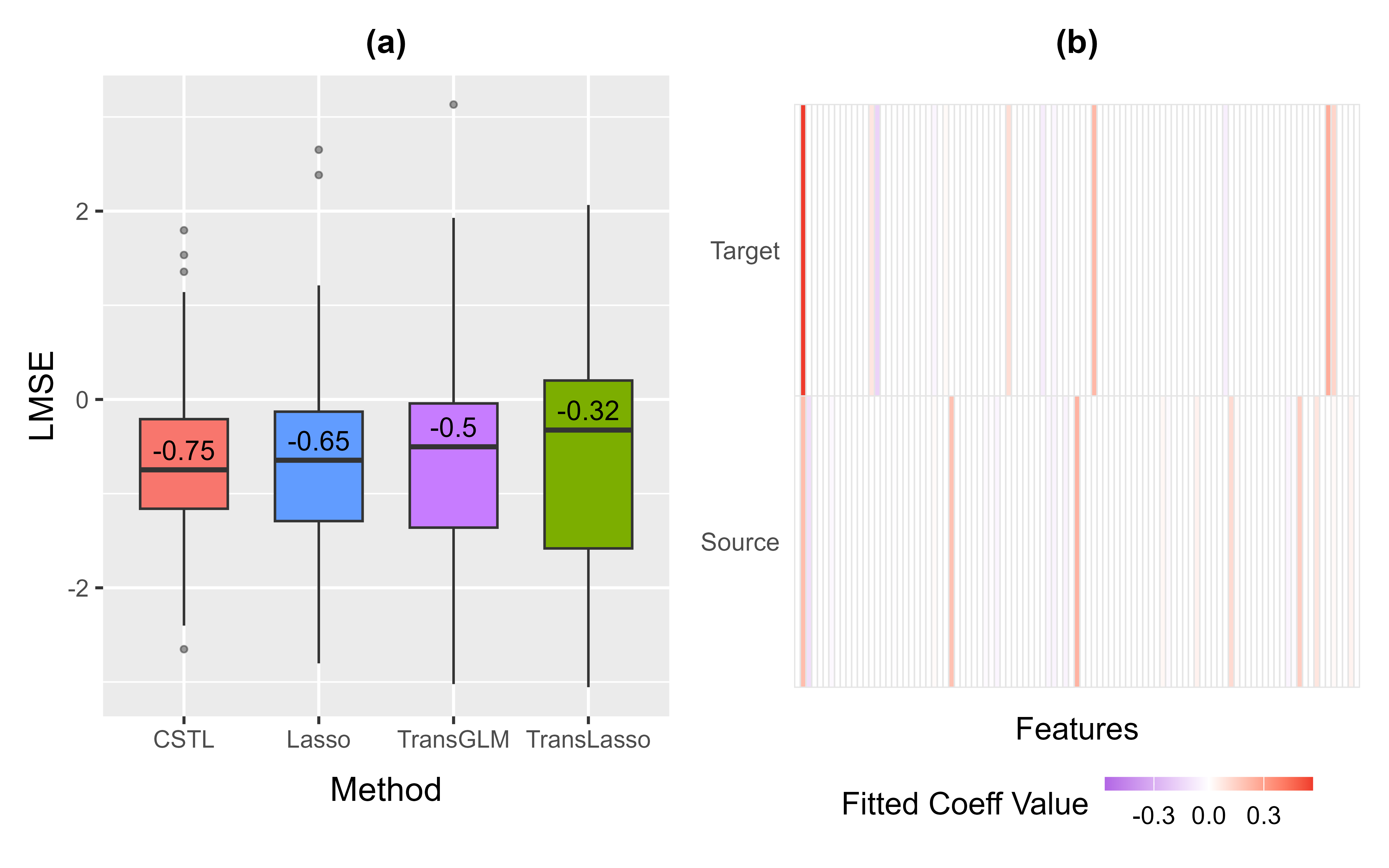}
	\caption{(a) Logarithm of  mean squared prediction errors (LMSE) over 100 repetitions on the Communities and Crime dataset. (b) Coefficients estimated by Lasso on source and target data.}
	\label{fig:realdata}
\end{figure}

We evaluate our method using the Communities and Crime Unnormalized dataset, which comprises community-level statistics from various U.S. regions and is publicly available from the UCI Machine Learning Repository.
Following the experimental setup of \citet{liu2025robust}, we formulate a linear regression task to predict the rate of violent crimes per 100,000 population, employing 99 demographic attributes as predictors. Domains are defined based on U.S. states: New Jersey (NJ), with 211 samples, serves as the source domain, while Washington (WA), with 40 samples, constitutes the target domain. For each replication, the target data are randomly split into an 80\% training set and a 20\% holdout testing set, with all source data used for training. To ensure statistical stability, this random splitting is repeated 100 times. The logarithm of the mean squared prediction error (LMSE) across these replications is reported in Figure~\ref{fig:realdata}(a).

As evidenced in Figure~\ref{fig:realdata}(a), CSTL achieves superior predictive performance, significantly outperforming all benchmarks. Notably, TransLasso and TransGLM perform worse than the Lasso baseline, indicating a clear case of negative transfer. This outcome finds a compelling explanation in Figure~\ref{fig:realdata}(b), which compares the coefficient estimates from separate Lasso fits on the source and target domains. While some coefficients share similar values, they are associated with semantically different covariates, illustrating a scenario of cross-semantic signal similarity. This phenomenon disrupts conventional transfer learning methods like TransLasso and TransGLM, which rely on strict covariate alignment for identifying transferable signals. CSTL, by design, overcomes this fundamental limitation by directly assessing signal transferability without requiring semantic correspondence. This real-world analysis provides strong empirical validation for our method and underscores the practical relevance of the challenges identified in our simulation studies.

\section*{Supplementary Material}
To save space, all technical proofs of theorems are included in the online supplementary material.

\vspace{-1em}
\section*{Declarations}
The authors do not have any conflict of interest to declare.

\bibliographystyle{apalike}
\bibliography{refs}

\end{document}